\documentclass[compsoc,conference,a4paper,10pt,times]{IEEEtran}
\IEEEoverridecommandlockouts
\usepackage[numbers]{natbib}
\usepackage{amsmath,amssymb,amsfonts}
\usepackage{nicefrac}
\usepackage{algpseudocode}
\usepackage{algorithm}
\usepackage{threeparttable}
\usepackage{etoolbox}
\usepackage{booktabs}
\usepackage[nolist]{acronym}
\usepackage[disable,colorinlistoftodos]{todonotes}
\usepackage{siunitx}
\usepackage{multirow}
\usepackage{multicol}
\usepackage{pifont}
\usepackage{soul}
\usepackage{caption}
\usepackage{subcaption}
\usepackage{graphicx}
\usepackage{titlesec}
\titleformat{\paragraph}[runin]
  {\normalfont\bfseries}      
  {\theparagraph}{1em}{}      
\titlespacing*{\paragraph}{0pt}{1.25ex plus .5ex minus .2ex}{1em}

\usepackage[colorlinks=true,urlcolor=black]{hyperref}
\usepackage[nameinlink,english,capitalise]{cleveref}
\def\BibTeX{{\rm B\kern-.05em{\sc i\kern-.025em b}\kern-.08em
    T\kern-.1667em\lower.7ex\hbox{E}\kern-.125emX}}

\makeatletter
\AtBeginDocument
 {
   \def\ltx@label#1{\cref@label{#1}}
   \def\label@in@display@noarg#1{\cref@old@label@in@display{#1}}
\def\label@in@mmeasure@noarg#1{%
    \begingroup%
      \measuring@false%
      \cref@old@label@in@display{#1}
    \endgroup}%
 } %
\makeatother

\makeatletter
\newcommand{\acposs}[1]{%
 \expandafter\ifx\csname AC@\AC@prefix#1\endcsname\AC@used
   \acs{#1}'s%
 \else
   \aclu{#1}'s (\acs{#1})%
 \fi
}
\makeatother

\newcommand{\fo}[2]{\texthl{#1}\todo[color=olive, inline]{FO: #2}}

\newcommand{\silvan}[2]{\texthl{#1}\todo[color=tracegreen, inline]{Silvan: #2}}

\let\oldFootnote\footnote
\newcommand\nextToken\relax

\renewcommand\footnote[1]{%
    \oldFootnote{#1}\futurelet\nextToken\isFootnote}

\newcommand\isFootnote{%
    \ifx\footnote\nextToken\textsuperscript{,}\fi}

\definecolor{tracered}{RGB}{213, 94, 0}
\definecolor{tracegreen}{RGB}{00, 158, 115}

\newtoggle{anonymous}
\togglefalse{anonymous} 

\begin{document}

\acrodef{aes}[AES]{advanced encryption standard}\acused{aes}
\acrodef{adb}[ADB]{Android Debug Bridge}
\acrodef{alu}[ALU]{arithmetic logic unit}\acused{alu}
\acrodef{arf}[ARF]{architecture and reference framework}
\acrodef{ascad}[ASCAD]{ANSSI SCA database}
\acrodef{avavan5}[AVA\_VAN.5]{Vulnerability Analysis Level 5}\acused{avavan5}

\acrodef{bfu}[BFU]{before first unlock}

\acrodef{cc}[CC]{Common Criteria}
\acrodef{cnn}[CNN]{convolutional neural network}
\acrodef{cpu}[CPU]{central processing unit}\acused{cpu}
\acrodef{cpa}[CPA]{correlation power analysis}

\acrodef{dram}[DRAM]{dynamic random-access memory}\acused{dram}
\acrodef{dsa}[DSA]{digital signature algorithm}
\acrodef{dpa}[DPA]{differential power analysis}

\acrodef{eal4+}[EAL4+]{evaluation assurance level 4}\acused{eal4+}
\acrodef{eal2+}[EAL2+]{evaluation assurance level 2 augmented}\acused{eal2+}
\acrodef{ecc}[ECC]{elliptic curve cryptography}
\acrodef{ecdh}[ECDH]{elliptic curve Diffie-Hellman}\acused{ecdh}
\acrodef{ecdsa}[ECDSA]{elliptic curve digital signature algorithm}
\acrodef{eddsa}[EdDSA]{Edwards curve digital signature algorithm}
\acrodef{em}[EM]{electromagnetic}
\acrodef{emfi}[EMFI]{electromagnetic fault injection}
\acrodef{eu}[EU]{European Union}
\acrodef{eudi}[EUDI]{European Digital Identity}

\acrodef{fi}[FI]{fault injection}

\acrodef{gpio}[GPIO]{general-purpose input/output}\acused{gpio}
\acrodef{gui}[GUI]{graphical user interface}
\acrodef{gpu}[GPU]{graphics processing unit}\acused{gpu}

\acrodef{hsm}[HSM]{hardware security module}
\acrodef{hnp}[HNP]{hidden number problem}

\acrodef{ic}[IC]{integrated circuit}
\acrodef{iot}[IoT]{internet of things}
\acrodef{ip}[IP]{intellectual property}
\acrodef{ir}[IR]{infrared}
\acrodef{id}[ID]{ID}\acused{id}
\acrodef{i2c}[I2C]{inter-integrated circuit}\acused{i2c}

\acrodef{lfi}[LFI]{laser fault injection}

\acrodef{mlp}[MLP]{multilayer perceptron}

\acrodef{naf}[NAF]{non-adjacent form}
\acrodef{nfc}[NFC]{near-field communication}\acused{nfc}
\acrodef{npu}[NPU]{neural processing unit}

\acrodef{os}[OS]{operating system}

\acrodef{pc}[PC]{personal computer}\acused{pc}
\acrodef{pcb}[PCB]{printed circuit board}\acused{pcb}
\acrodef{pin}[PIN]{personal identification number}\acused{pin}
\acrodef{pop}[PoP]{package-on-package}
\acrodef{pwm}[PWM]{pulse width modulation}

\acrodef{rsa}[RSA]{Rivest-Shamir-Adleman}\acused{rsa}

\acrodef{sca}[SCA]{side-channel analysis}\acrodefplural{sca}[SCA]{side-channel analyses}
\acrodef{sdr}[SDR]{software-defined radio}
\acrodef{se}[SE]{secure element}
\acrodef{sid}[SID]{intrinsic system identifier}
\acrodef{sim}[SIM]{subscriber identity module}\acused{sim}
\acrodef{sram}[SRAM]{static random-access memory}
\acrodef{soc}[SoC]{system on chip}
\acrodef{stft}[STFT]{Short-Time Fourier Transformation}
\acrodef{ssh}[SSH]{secure shell protocol}\acused{ssh}
\acrodef{spa}[SPA]{simple power analysis}

\acrodef{tee}[TEE]{trusted execution environment}
\acrodef{tls}[TLS]{transport layer security}\acused{tls}
\acrodef{toe}[TOE]{target of evaluation}
\acrodef{tvla}[TVLA]{test vector leakage assessment}

\acrodef{uv}[UV]{ultraviolet}
\acrodef{usvp}[uSVP]{unique shortest vector problem}
\acrodef{usb}[USB]{universal serial bus}\acused{usb}
\acrodef{usb-c}[USB-C]{universal serial bus type-c}\acused{usb-c}

\acrodef{wscd}[WSCD]{wallet-secure cryptographic device}
\acrodef{wifi}[Wi-Fi]{Wi-Fi}\acused{wifi}
\acrodef{wnaf}[wNAF]{w-ary non-adjacent form}

\title{Breaking ECDSA with Electromagnetic Side-Channel Attacks: Challenges and Practicality on Modern Smartphones}

\iftoggle{anonymous}{
}{
\author{
	\IEEEauthorblockN{
	Felix Oberhansl\IEEEauthorrefmark{1},
	Marc Schink\IEEEauthorrefmark{1}\IEEEauthorrefmark{3},
	Nisha Jacob Kabakci\IEEEauthorrefmark{1},\\
	Michael Gruber\IEEEauthorrefmark{1},
	Dominik Klein\IEEEauthorrefmark{2},
	Sven Freud\IEEEauthorrefmark{2},
	Tobias Damm\IEEEauthorrefmark{2},\\
	Michael Hartmeier\IEEEauthorrefmark{1},
	Ivan Gavrilan\IEEEauthorrefmark{1}\IEEEauthorrefmark{3},
	Silvan Streit\IEEEauthorrefmark{1}\IEEEauthorrefmark{3},
	Jonas Stappenbeck\IEEEauthorrefmark{1},
	Andreas Zankl\IEEEauthorrefmark{1}\IEEEauthorrefmark{3}
	}
	\IEEEauthorblockA{\IEEEauthorrefmark{1}Fraunhofer Institute for Applied and Integrated Security (AISEC), Garching, Germany\\
	Email: \{felix.oberhansl, marc.schink, nisha.jacob, michael.gruber,\\ michael.hartmeier, ivan.gavrilan, silvan.streit, andreas.zankl\}@aisec.fraunhofer.de\\
	Email (Jonas Stappenbeck): jonas.stappenbeck@tum.de}
	\IEEEauthorblockA{\IEEEauthorrefmark{2}German Federal Office for Information Security (BSI), Bonn, Germany\\
	Email: \{dominik.klein, sven.freud, tobias.damm\}@bsi.bund.de}
	\IEEEauthorblockA{\IEEEauthorrefmark{3}Technical University of Munich (TUM), Munich, Germany}
}
}

\maketitle

\begin{abstract}
Smartphones handle sensitive tasks such as messaging and payment and may soon support critical electronic identification through initiatives such as the \ac{eudi} wallet, currently under development.
Yet the susceptibility of modern smartphones to physical \ac{sca} is underexplored, with recent work limited to pre-2019 hardware.
Since then, smartphone \ac{soc} platforms have grown more complex, with heterogeneous processor clusters, sub \SI{10}{\nano\meter} nodes, and frequencies over \SI{2}{\giga\hertz}, potentially complicating \ac{sca}.
In this paper, we assess the feasibility of \ac{em} \ac{sca} on a Raspberry Pi~4, featuring a Broadcom~BCM2711 and a Fairphone~4 featuring a Snapdragon~750G~5G \ac{soc}.
Using new attack methodologies tailored to modern \acp{soc}, we recover \acs{ecdsa} secrets from OpenSSL by mounting the Nonce@Once attack of Alam et al. (Euro S\&P 2021) and show that a libgcrypt countermeasure does not fully mitigate it.
We present case studies illustrating how hardware and software stacks impact \ac{em} \ac{sca} feasibility.
Motivated by use cases involving security-critical mobile applications such as the currently developed \ac{eudi} wallet, we survey Android cryptographic implementations and define representative threat models to assess the attack.
Our findings show weaknesses in \acs{ecdsa} software implementations and underscore the need for independently certified \acp{se} in all smartphones.
\end{abstract}



\section{Introduction}
\label{sec:intro}

In today's world, smartphones have become indispensable in everyday life, playing crucial roles in messaging, payments, and identification, among other sensitive tasks.
Their significance continues to grow, highlighted by the \acposs{eu} plan to introduce a standardized \ac{eudi} wallet by 2026~\cite{eidas}.
This demonstrates that smartphones are valuable targets for attackers, and no attack path should be considered irrelevant.
Although software exploits, such as Operation Triangulation~\cite{operation_triangulation}, are usually the primary concern due to their less stringent threat model, the lack of studies on physical \ac{sca} on modern smartphones calls for more exploration in this area.
Furthermore, the usage of smartphones for financial transactions, health monitoring, and digital identities (\ac{eudi}) warrants the investigation of more powerful adversarial models.
Since existing hardware cannot be changed, countermeasures to newly developed hardware attacks must be implemented in software, which is often challenging and insufficient to address the underlying issues.

Physical attacks are directed at the implementation of an algorithm and its physical effects, in our case the \ac{em} emissions from smartphones.
Such attacks can be advantageous if, for example, cryptographic operations are conducted within a \ac{tee} which offers software isolation but does not protect against hardware attacks.
In this paper, we show how to recover the secret \ac{ecdsa} nonce values from software libraries executed on a modern smartphone by \ac{em} \ac{sca}.

\paragraph{Attack scope and threat model}
Our attack scope is to investigate general-purpose application-class processors and popular cryptographic software libraries such as OpenSSL\footnote{\url{https://github.com/openssl/openssl}} and libgcrypt\footnote{\url{https://www.gnupg.org/software/libgcrypt/index.html}} that may run inside or outside the context of a \ac{tee}.
We do not investigate tamper-resistant hardware such as \acp{se}.
While some flagship and selected mid‑range devices include such hardware, many devices rely solely on the \ac{tee}.
This is particularly relevant as phones are increasingly used for sensitive use cases, leaving devices without an \ac{se} to store crucial secrets outside tamper‑resistant hardware.

Executing such an attack comes with certain prerequisites: We consider a local, passive \ac{em} side‑channel adversary with physical access to the smartphone 
and lab‑grade measurement equipment.
In order to assess the impact and the limitations of our result for practical smartphone security, we discuss two threat models with respect to these prerequisites in \cref{sec:threatmodel}.
In one, the device is screen-locked and the attacker cannot bypass this lock.
In the second model, the attacker can unlock the device, representing (temporary) device theft in combination with weak screen-lock mechanisms.
We take security-critical mobile applications, such as the \ac{eudi} wallet and its ongoing development as an example throughout this paper in order to discuss the 
potential real-world impact of our findings.

\paragraph{Related work}
The practicality of \ac{sca} on a smartphone's application processor is hard to judge from state-of-the-art literature.
Given the advancements in \ac{soc} technologies and smartphones, it is noteworthy that devices in recent works are more than five years old (\cref{tab:sota}).
The authors of~\cite{alam_nonce_at_once, alam_one_and_done} demonstrated successful \ac{em} attacks on \ac{rsa} and \ac{ecc} on smartphones.
The Nonce@Once attack~\cite{alam_nonce_at_once} recovered \ac{ecdsa} nonces from conditional swap operations.
In~\cite{genkin_ecdsa_mobile_nonintrusive, belgarric_ecdsa_android}, \ac{naf} and \ac{wnaf} \ac{ecdsa} implementations were investigated on multiple smartphones.
In~\cite{lisovets_iphone_bootloader}, the authors extracted a device unique key from the iPhone~4's \ac{aes} engine.
The authors of~\cite{vasselle_bootloader} recovered the \ac{aes} firmware encryption key of the first stage bootloader on an undisclosed mobile device.
Apple's CoreCrypto Library and the ARMv8 Cryptographic Extensions for \ac{aes} were analyzed on an iPhone~7 in \cite{haas_apple_vs_ema}.
Challenges when targeting \ac{aes} with \ac{em} \ac{sca} on an ARM Cortex-A architecture were described in~\cite{bhasin_aes_arm_cortex_a}.
They used a Raspberry Pi 4 for their experiments.
For \ac{em} attacks on older devices we refer the reader to~\cite{longo_soc_it_to_em, genkin_ecdsa_mobile_nonintrusive, goller_sca_smartphones_standard_radio} and the survey in \cite{shepherd_physical_attacks_mobile_survey}.
More recent smartphones, such as the iPhone 8 Plus (2017)~\cite{cronin_charger_surfing}, Xiaomi Redmi 7A (2019)~\cite{sayakkara_forensic_em}, iPhone 13 and 14 Pro (2022)~\cite{navanesan_multiple_cpu_forensic}, or Google Pixel 9 (2024) and Samsung Galaxy S25 (2025)~\cite{wang_pixnapping} were only used for forensic \acp{sca}, where no cryptographic algorithms were targeted; instead, \ac{pin} entry~\cite{cronin_charger_surfing}, browsing activity~\cite{sayakkara_forensic_em}, differentiation between software activities~\cite{navanesan_multiple_cpu_forensic} or display contents~\cite{wang_pixnapping} were investigated.

\paragraph{Complexity of contemporary smartphones}
Nowadays, complex \ac{soc} platforms are at the heart of smartphones.
This complexity impacts the feasibility of \ac{sca}.
Modern mobile \acp{soc} feature multiple \acp{cpu}, often in clusters of two or more.
These clusters are optimized either for power consumption, performance, or a trade-off between the two.
This allows versatile scheduling options, resulting in a vast evaluation space for \ac{em} \ac{sca}.
The \ac{cpu} cores can dynamically adapt their clock frequency with the upper limit reaching more than \SI{3}{\giga\hertz}.
The \ac{ic} for the smartphone is usually fabricated at advanced semiconductor nodes, such as \SI{10}{\nano\meter} and below. 
All this potentially increases the challenge for successful \ac{em} side-channel measurements.

\begin{table*}
        \centering
        \begin{threeparttable}
		\caption{Smartphones in state-of-the-art literature that were investigated using \ac{sca}.}
                \label{tab:sota}
                \begin{tabular}{lclllll}
                        \toprule
                        Device                     & Release             & Reference                                    & \ac{soc}                          & Arch.                 & \ac{cpu}\tnote{a}                    & Node                                \\
                        \midrule
                        iPhone 4                   & 2010                & \cite{lisovets_iphone_bootloader}            & Apple A4                          & n/a\tnote{c}          & n/a\tnote{c}                        & \SI{45}{\nano\meter}                \\
                        iPhone 4                   & 2010                & \cite{genkin_ecdsa_mobile_nonintrusive}      & Apple A4                          & 32-bit                & 1xA8 @ \SI{800}{\mega\hertz}         & \SI{45}{\nano\meter}                \\
                        Xperia X10                 & 2010                & \cite{genkin_ecdsa_mobile_nonintrusive}      & Snapdragon S1                     & 32-bit                & 1xA5 @ \SI{800}{\mega\hertz}         & \SI{45}{\nano\meter}                \\
                        -                          & 2012                & \cite{belgarric_ecdsa_android}               & Snapdragon S4                     & 32-bit                & 2xA5 @ \SI{1.2}{\giga\hertz}         & \SI{45}{\nano\meter}                \\
                        -\tnote{b}                 & 2012                & \cite{vasselle_bootloader}                   & -\tnote{b}                        & n/a\tnote{c}          & n/a\tnote{c}                        & \SI{32}{\nano\meter}                \\
                        Galaxy Centura             & 2013                & \cite{alam_one_and_done}                     & Snapdragon S1                     & 32-bit                & 1xA5 @ \SI{800}{\mega\hertz}         & \SI{45}{\nano\meter}                \\
                        \multirow{2}*{iPhone 7}    & \multirow{2}*{2016} & \multirow{2}*{\cite{haas_apple_vs_ema}}      & \multirow{2}*{Apple A10 Fusion}   & \multirow{2}*{64-bit} & 2xHurricane @ \SI{2.34}{\giga\hertz} & \multirow{2}*{\SI{14}{\nano\meter}} \\
                                                   &                     &                                              &                                   &                       & 2xZephyr @ \SI{1.05}{\giga\hertz}    &                                     \\
                        Alcatel Ideal              & 2016                & \cite{alam_nonce_at_once, alam_one_and_done} & Snapdragon 210                    & 32-bit                & 4xA7 @ \SI{1.1}{\giga\hertz}         & \SI{28}{\nano\meter}                \\
                        ZTE ZFIVE                  & 2018                & \cite{alam_nonce_at_once}                    & Snapdragon 425                    & 64-bit                & 4xA53 @ \SI{1.4}{\giga\hertz}        & \SI{28}{\nano\meter}                \\
                        \midrule
                        Raspberry Pi 4             & 2019                & This Work, \cite{bhasin_aes_arm_cortex_a}    & BCM2711                           & 64-bit                & 4xA72 @ \SI{1.8}{\giga\hertz}        & \SI{28}{\nano\meter}                \\
                        \midrule
                        \multirow{2}*{Fairphone 4} & \multirow{2}*{2021} & \multirow{2}*{This Work}                     & \multirow{2}*{Snapdragon 750G 5G} & \multirow{2}*{64-bit} & 2xA77 @ \SI{2.21}{\giga\hertz}       & \multirow{2}*{\SI{8}{\nano\meter}}  \\
                                                   &                     &                                              &                                   &                       & 6xA55 @ \SI{1.8}{\giga\hertz}        &                                     \\
                        \bottomrule
                \end{tabular}
                \begin{tablenotes}
                        \item[a]{All \acp{cpu} are ARM Cortex cores.}
                        \item[b]{Device not disclosed in paper, release year and semiconductor node were obtained from authors.}
                        \item[c]{The attacks do not target the \ac{cpu} but an internal \ac{aes} engine.}
                \end{tablenotes}
        \end{threeparttable}
\end{table*}

\paragraph{Contribution}
Our main contribution is a thorough analysis of the practicality of \ac{em} \ac{sca} on a contemporary smartphone.
\cref{tab:sota} highlights the gap between smartphones that were targeted in existing literature and the Fairphone~4, our main target device.
More precisely, we make contributions in the following areas:
\begin{itemize}
	\item \textbf{Threat models and vulnerability assessment.} We define realistic threat models and provide an overview of how cryptography is used on Android smartphones. We later assess our practical attack findings in the context of these defined threat models.
	\item \textbf{Methodology.} Starting from the attacks on \ac{ecc} conditional swap operations in~\cite{nascimento_ecc_cmov_sca,nascimento_ecc_embedded_sca,alam_nonce_at_once} we adapt the attack methodology to reproduce it on more complex devices. Among others, we investigate activity-modulated signals at different frequency bands and methods for leakage assessment and exploitation. We also analyze the countermeasures proposed in \cite{alam_nonce_at_once} and implemented in libgcrypt and come to the conclusion that they offer insufficient protection.
	\item \textbf{Hardware complexity.} We start our investigations with a Raspberry Pi 4 as baseline to evaluate our \ac{sca} methodology. Subsequently we investigate the Fairphone~4's Snapdragon 750G 5G \ac{soc} which is also used in phones such as the Samsung Galaxy A52 5G and the Xiaomi Mi 10T Lite. We show the impact that multi-cluster \acp{cpu} with high frequencies and dynamic frequency scaling have on the attack.
	\item \textbf{Software complexity.} We evaluate the complexity introduced by the software stack in case studies on the Fairphone~4, where we can install both Android and a native Linux \ac{os} to which we can easily apply constraints. To the best of our knowledge, we are the first to investigate complete Android apps and the increased complexity this introduces. We demonstrate that the attack is possible despite dynamic scheduling, dynamic frequency scaling and background activities.
\end{itemize}

Our findings suggest that, while \acp{sca} of smartphones is no longer as simple as in~\cite{alam_nonce_at_once, alam_one_and_done, genkin_ecdsa_mobile_nonintrusive, belgarric_ecdsa_android}, attacks are still possible and therefore remain a threat to contemporary smartphones.
Thus, the usage of certified secure elements should be mandatory for critical applications.

Code and traces that highlight our methodology are publicly available\footnote{\url{https://github.com/Fraunhofer-AISEC/breaking-ecdsa-emsca-smartphones}}.
Following coordinated disclosure, we made our findings available to the affected libraries (\cref{appendix:disclosure}).

\section{Background}
\label{sec:background}

In this section, we introduce general preliminaries on \ac{em} \ac{sca}, \ac{ecc} and \ac{ecdsa}.

\subsection{\Acl{em} \acl{sca}}

\ac{sca} exploits unintentional leakage (e.g., timing~\cite{kocher_sca}, power consumption~\cite{kocher_dpa}, or electromagnetic emissions~\cite{agrawal_em_sca}) from a system to infer sensitive data.
Due to their accessibility, smart cards and other embedded devices are the most investigated targets for these physical side channels.
In the past, more complex devices such as \acp{pc} or smartphones were mostly investigated for logical, i.e., microarchitectural, side channels such as Spectre~\cite{kocher_spectre}.
Only recently, the investigation of such devices with regard to \ac{em} \ac{sca} gained attention~\cite{genkin_ecdsa_mobile_nonintrusive,alam_one_and_done,alam_nonce_at_once,belgarric_ecdsa_android,genkin_ecdh_sca_pc}.

Profiled \ac{sca} is a special case of \ac{sca}, which relies on collecting measurement data from a device similar or identical to the target in order to train an attack model during the profiling phase.
The adversary requires access to such a device and the ability to run and monitor the cryptographic operations for profiling.
In the attack phase, profiled attacks typically require only a small number of traces, making them highly efficient due to the pre-built profiles.

\subsection{Electronic identification and \Ac{ecdsa}}
\label{subsec:ecdsa}

\Acp{dsa} ensure the integrity and authenticity of digital transactions.
\ac{ecdsa} uses elliptic curves as a mathematical primitive.
It is favored in many applications due to its ability to provide strong security with small key sizes, making it efficient in terms of computational and storage requirements.

\Acp{dsa} rely on private–public key pairs.
The public key is shared with a remote party.
The secret, private key remains with the signer.
A common use case is electronic identification to a relying party.
For example, if Alice wants to authenticate to Bob, he generates a challenge and sends it to Alice.
Alice signs the challenge and returns the signature.
Bob verifies the signature using Alice's public key.

For an \ac{ecdsa} signature to be created, first an elliptic curve in form of its underlying field and equation must be chosen.
Each curve has a base point $G$ which generates a cyclic group of order $n$.
The curve's point at infinity, also called neutral element, $\mathcal{O}$ is $\mathcal{O} = nG$.
Points can be represented in projective coordinates $(X : Y : Z)$ or affine coordinates $(x, y)$.

To sign a message (e.g., challenge), Alice selects a random scalar $d_A$ as her secret private key and the curve point $Q_A$, obtained by the scalar-by-point multiplication $Q_A=d_AG$, as her public key.
A message $m$ is signed by first getting its digest $z$ from a cryptographic hash function.
Subsequently, the signer chooses a random nonce $k$ such that $1 \leq k \leq n-1$.
The nonce is used as a scalar for a scalar-by-point multiplication that yields the coordinates $x$ and $y$ of a new point on the curve: $(x,y) = kG$.
The signature is the tuple $(r,s)$, where $r = x \pmod n$ and $s = k^{-1}(z + r \cdot d_A) \pmod n$.
The process is repeated for a new random $k$ if either $r$ or $s$ are zero.

To verify the signature, Bob can compute $z$ from the message $m$ and $w = s^{-1} \pmod n$, $u_1 = z \cdot w \pmod n$ and $u_2 = r \cdot w \pmod n$.
From this, the candidate coordinates $(x',y')$ can be obtained as $(x',y') = u_1G + u_2Q_A$.
If $x' \equiv r \pmod n$, Bob accepts the signature.

The nonce $k$ must be secret, random and unique for each new signature that is created.
If an adversary knows $k$, she can recover $d_A$ from the signature $(r,s)$.
We summarize relevant works on \ac{sca} of \ac{ecdsa} in \cref{sec:vulnerability}.


\section{Threat models}
\label{sec:threatmodel}

To evaluate the practicality of our attack and clearly explain the motivation for hardware attacks on smartphones, we introduce two threat models.
These models serve primarily to assess how well our analyses, presented in \cref{sec:contemporarysmartphones}, align with realistic attacker capabilities in \cref{sec:impact}.
We distinguish threat models based on whether the adversary can unlock the phone (\cref{subsec:scaunlocked}) or not (\cref{subsec:scalocked}).

\subsection{\ac{em} \acp{sca} on unlocked smartphone}
\label{subsec:scaunlocked}

In this subsection, we assume that the adversary has an unlocked phone in her possession and user-level access.
In this case it might seem questionable whether \ac{sca} is needed.
The adversary can already access the vast majority of the victim's data and install malicious software.
However, certain cryptographic secrets are not accessible, e.g., secrets stored within a \ac{tee} or \ac{se}.
While the \ac{se} typically features hardware countermeasures against physical \ac{sca}, the \ac{tee} does not, as it uses the same processor as generic software running on the device.

The unlocked phone might allow an adversary to trigger cryptographic processes, as long as no further authentication besides the unlock mechanism is required (see authorization for key usage below).
\ac{sca} can then be used to recover the involved cryptographic secrets.
Depending on how invasive the preparations needed for the \ac{em} \ac{sca} are, the device could afterwards be returned to its owner without raising suspicion.

To underline the realism of this threat model, we emphasize that it is equivalent to scenarios where the adversary obtains a screen-locked smartphone, but can easily bypass the lock.
Recent work has shown that the security of a simple unlock \ac{pin} depends on whether countermeasures such as throttling and blocklisting are deployed and designed reasonably~\cite{markert_unlock_pins}.
Graphical passwords such as unlock patterns are often biased and offer less security than guessing a three-digit \ac{pin}~\cite{uellenbeck_unlock_patterns}.
Additionally, they face the threat of smudge attacks~\cite{aviv_smudge_attacks}, which might apply to \acp{pin} as well.
Biometric mechanisms such as a fingerprint sensor can be overcome as shown in~\cite{chen_infinity_gauntlet, telleria_attack_potential_fingerprint, tao_recovering_fingerprints}.
Also, facial recognition systems for unlocking a phone are potentially vulnerable to presentation attacks~\cite{zheng_spoofing_fa}.
The practical feasibility of bypassing a screen lock depends on the specific locking mechanism, the user’s vigilance, and the resources an attacker is willing to expend.

It is important to note that Android allows app developers to additionally gate access to cryptographic keys\footnote{\url{https://source.android.com/docs/security/features/authentication}}.
For example, a phone could be unlocked with a \ac{pin}, but usage of a cryptographic key could require additional biometric authentication.
Our threat model here assumes the adversary can bypass device unlock and any key-use authorization mechanisms.

Note that for digital signing with non-repudiation, the adversary can also be the owner:
Non-repudiation implies that the private key has sufficient protections in place that the owner cannot later repudiate actions she takes with the key.
If a private key is compromised, it undermines non-repudiation.

\subsection{\ac{em} \acp{sca} on locked smartphone}
\label{subsec:scalocked}

If the phone is insurmountably locked, the adversary usually has no possibility to read out data directly.
We assume that no pairing via \ac{usb}, Bluetooth, or \ac{adb} is possible.
Therefore \ac{sca} of cryptographic software running both inside and outside the context of a \ac{tee} is a reasonable path of attack.
\Acp{sca} on locked smartphones can be further divided into two scenarios.
A scenario where the device is (temporarily) stolen and one where the owner always remains in possession of the device.

If the owner remains in possession of the smartphone, attacks usually must not require any noticeable physical modifications to the device and must be carried out within a short period of time.
The adversary could use a probe placed on the display or the device's casing (e.g., by hiding it in a desk), as in \cite{alam_nonce_at_once, alam_one_and_done, belgarric_ecdsa_android, genkin_ecdsa_mobile_nonintrusive} or hidden in a charger~\cite{genkin_ecdsa_mobile_nonintrusive}.

If the device is stolen, the applicability of attacks is reduced to scenarios, where cryptographic operations are triggered by normal system activity (e.g., decryption of incoming messages or \ac{tls} handshakes).
This might require notifications to be enabled for the locked phone.
The Android auto-reboot feature~\cite{google_auto_reboot} could complicate this threat model.
Devices for which this feature is enabled and which are not unlocked for three days are automatically rebooted.
After the reboot, disk encryption might be active and the device is set into the \ac{bfu} state, which might disable certain unlock methods and features such as the automatic decryption of incoming messages.
However, this does not mitigate the attack, as the targeted cryptographic routine is likely to be triggered at least once in this time span.
For messaging, an adversary could send a message to the victim to induce execution.

\section{Cryptography on Android: Implementations and attack surfaces}
\label{sec:vulnerability}

In this section, we dive deeper into the potential vulnerability of smartphones to \ac{em} \ac{sca} by investigating how cryptography can be implemented on Android smartphones.
We review considerations for the \ac{eudi} wallet and provide an overview of \ac{sca} targeting \ac{ecdsa}.

\subsection{Cryptography on Android smartphones}
\label{subsec:vulnerability:devs}

We focus our investigation on Android smartphones due to the transparency and openness of the Android ecosystem and its wide deployment.
We emphasize that this openness does not imply less security compared to other platforms.
The Android Keystore~\cite{android_keystore} is the default provider for cryptography on Android smartphones.
For compatibility and the support for algorithms not included in the Android Keystore, most phones also have alternative providers such as OpenSSL or Bouncy Castle\footnote{\url{https://www.bouncycastle.org/}}.
The Android Keystore is the only possibility to use a \ac{tee} or StrongBox, the latter being Android's term for an \ac{se}, i.e., dedicated tamper-resistant hardware.

The Android Device Security Database\footnote{\url{https://www.android-device-security.org}} provides an overview of security features available on phones, e.g., which devices include StrongBox and which do not~\cite{leierzopf_adsdb}.
High-end devices such as Google Pixel phones and the Samsung Galaxy S series support StrongBox.
However, even if StrongBox is available, there are important caveats, which might block its wide adoption for applications.
For one, \ac{tee} is the general default for the Android Keystore and developers are urged to only use StrongBox for applications requiring the highest level of security~\cite{android_keystore}.
Since StrongBox is typically implemented outside of the \ac{soc}, within a dedicated \ac{ic}, operations are much slower and more limited with regards to concurrency than if the \ac{tee} were used~\cite{android_keystore}.
Therefore, an app must explicitly request StrongBox-backed keys if it shall be used over the \ac{tee}.
Popular messaging apps like Signal\footnote{\url{https://github.com/signalapp/Signal-Android}}, Telegram\footnote{\url{https://github.com/DrKLO/Telegram}} and Threema\footnote{\url{https://github.com/threema-ch/threema-android}} do not do this, since the performance requirements for messaging are demanding with regard to the amount of data that needs to be encrypted.
Second, not all algorithms are necessarily supported by StrongBox.
We read out the supported algorithms on a Google Pixel~8, which features the Titan M2 \ac{se}~\cite{google_pixel_security}.
Of the \ac{ecc} signature algorithms, only the \textit{secp256r1} curve is supported by StrongBox.

Furthermore, not all modern smartphones are equipped with the StrongBox feature.
Even recent mid-range devices such as the Samsung Galaxy A55 or the Sony Xperia 10 do not include suitable hardware for StrongBox yet.
Of the twelve devices with a release after January 2022 that are listed in the database of~\cite{leierzopf_adsdb}, only five support StrongBox, while all of them support \ac{tee}.
In fact, all devices in the database dating back to January 2018 support \ac{tee}.
Thus, many smartphones will continue to rely on software-implemented cryptography on non‑tamper‑resistant hardware in the near future.


%

Security-critical mobile applications must store sensitive data, e.g., keys for electronic identification (\cref{sec:background}).
We use the \ac{eudi} wallet, which is currently being developed, to highlight the dilemma that arises when devices without secure hardware must store highly sensitive data.
Electronic identification is often built around \acp{dsa} such as \ac{ecdsa} (\cref{subsec:ecdsa}).
Within the wallet, the \ac{wscd} is the module in charge of managing the secret wallet keys for its holder.
The \ac{eu} requires all \acp{wscd} to be tamper-resistant and certified.
The draft of the implementation act \cite{eidas_implementing_regulation} lists the \ac{cc} assurance level \ac{eal4+} and the \ac{avavan5} vulnerability assessment~\cite{cc_eal} as potential certification targets.
This demonstrates a strong commitment to the wallet's resilience, also with regard to hardware attacks.
Furthermore, it underlines the necessity to investigate the susceptibility of smartphones to hardware attacks.

While there are smartphones that could potentially fulfill such a requirement (e.g., Google Pixel phones which include a Titan M2~\cite{google_pixel_security,trustcb_titan_certificate}), the \ac{eudi} wallet also plans to consider older smartphones and cheaper models lacking secure hardware~\cite{eidas_arf}.
According to the \ac{eudi} documentation, such devices shall rely on a remote \ac{hsm} to act as \ac{wscd}~\cite{eidas_arf}.
In this case, the wallet keys are protected, but the wallet app must authenticate itself to the remote \ac{hsm}, and the secret authentication key must be securely stored on a smartphone lacking secure hardware.

Without StrongBox, \ac{tee}-based implementations are the next best option for the \ac{eudi} wallet app.
\acp{tee} cannot be considered secure against hardware attacks, i.e., tamper-resistant.
For example, the Qualcomm \ac{tee} v5.8 on the Snapdragon 865 passed an \ac{eal2+} evaluation~\cite{tuev_qualcomm_tee}, which does not consider hardware attacks~\cite{cc_eal}.
Therefore, a \ac{tee}-based authentication of the wallet app to the remote \ac{wscd} would be an attractive target for \ac{em} \ac{sca} and our work aims to provide an objective quantification of the risks associated with such attacks.
\fo{}{Check if there is a certification target for this procedure. A slide of a workshop mentioned AVA\_VAN5 for complete WSCA once, but that would mean that certain smartphones could not use the EUDI at all?}

\subsection{\ac{sca} of ECC implementations}
\label{subsec:vulnerability:libs}

Within our work, we target open-source libraries and their \ac{ecdsa} implementations.
As discussed above, Android apps should use the Android Keystore for cryptography.
However, there are multiple reasons to focus on libraries such as OpenSSL and libgcrypt first.
For one, the implementation details of StrongBox and \ac{tee}, the default backends of the Keystore, are secret.
It therefore makes sense to start by examining a known implementation.
As far as \acp{tee} are concerned, the use of countermeasures is limited to software, as is the case in open-source libraries.
Second, libraries such as OpenSSL are also available as an alternative provider to the Keystore on most Android devices.
Finally, some apps ship their own libraries to ensure compatibility across devices or utilize specific features absent in the Keystore.
The Telegram Messenger, for example, relies on its own BoringSSL implementation~\cite{suvanka_telegram}.

The selection of \ac{ecdsa} as targeted signature algorithm is straightforward.
It is a mandatory algorithm for the Android Keystore and a likely choice for applications such as the \ac{eudi} wallet.

In this section, we review state-of-the-art side-channel attacks on \ac{ecc}, most notably attacks on the conditional swap operation~\cite{alam_nonce_at_once, nascimento_ecc_cmov_sca,nascimento_ecc_embedded_sca}.

Most \acp{sca} of \ac{ecc} focus on the scalar-by-point multiplication.
This is because it is a critical operation that involves either ephemeral keys, often referred to as nonces, that allow to recover the long-term keys, or long-term secrets directly (\cref{subsec:ecdsa}).




\begin{algorithm}
\caption{Double-and-always-add multiplication.}
\label{alg:doubleandadd}
\hspace*{\algorithmicindent} \textbf{Input:} $n$-bit scalar $k = {k_{n-1}, k_{n-2}, \dots, k_0}$; elliptic curve point $P$.\\
\hspace*{\algorithmicindent} \textbf{Output:} $R \gets [k]P$.
\begin{algorithmic}[1]
\State $R \gets \mathcal{O}$ \Comment{Neutral element}
\For{$i\gets n - 1$ \textbf{to} $0$}
  \State $R \gets 2R$
  \State $T \gets R + P$
	\State $R,T \gets \mathsf{CT\_SWAP}(R,T,k_i)$ \Comment{Swap if $k_i = 1$}
\EndFor
\end{algorithmic}
\end{algorithm}


\paragraph{Attacks on the conditional swap operation}
The double-and-always-add algorithm and the Montgomery ladder are two popular choices for implementing a constant-time scalar-by-point multiplication for Weierstrass and Edwards curves~\cite{nist_ecc}.
The former is shown in~\cref{alg:doubleandadd} and used by libraries such as libgcrypt.
OpenSSL uses a Montgomery ladder~\cite{montgomery_ladder}, which differs slightly from the procedure in \cref{alg:doubleandadd}.
The most relevant difference for this work is that the swap operation is either required twice per iteration of the for loop or the swap condition $cond$ is merged to $cond = k_i \oplus k_{i+1}$.
In both cases, the sequence of double-and-add operations is constant.
This is achieved by conditionally swapping $R$ and $T$.
The constant-time conditional swap operation~$\mathsf{CT\_SWAP}$ (\cref{alg:cswap}) can be exploited for \ac{sca}.
This was demonstrated previously on microcontrollers~\cite{nascimento_ecc_cmov_sca,nascimento_ecc_embedded_sca} and smartphones~\cite{alam_nonce_at_once}.
It is noteworthy that popular \ac{sca} countermeasures for \ac{ecc} such as coordinate re-randomization, point blinding and scalar blinding are not or only partially effective against the attack.
We discuss countermeasures in more detail in \cref{sec:revisiting} and \cref{sec:impact}.

\begin{algorithm}
	\caption{Constant-time conditional swap operation~$\mathsf{CT\_SWAP}$.}
\label{alg:cswap}
\hspace*{\algorithmicindent} \textbf{Input:} Two arrays $a,b$ of size $n$ machine words, representing a coordinate of an elliptic curve point; Swap condition $cond \in \{0, 1\}$.\\
\hspace*{\algorithmicindent} \textbf{Output:}\\If $cond = 1$: $a \gets b$, $b \gets a$;\\If $cond = 0$: $a \gets a$, $b \gets b$.
\begin{algorithmic}[1]
\State $mask \gets 0 - cond$ \Comment{$mask$ is all-ones or zeros}
\For{$i\gets0$ \textbf{to} $n-1$}
  \State $\delta \gets (a[i] \oplus b[i]) \; \& \; mask$
  \State $a[i] \gets a[i] \oplus \delta$
  \State $b[i] \gets b[i] \oplus \delta$
\EndFor
\end{algorithmic}
\end{algorithm}

The swap condition is related to a single bit of a secret scalar, which is either one, i.e., the points are swapped, or zero, i.e., the points are not swapped.
\cref{alg:cswap} shows how an arithmetic mask can be used to achieve a constant runtime independent of the swap condition.
The authors of \cite{alam_nonce_at_once} discussed that this comes at the cost of \ac{em} leakage amplification.
First, the single bit in $cond$ is expanded to either an all-zeros or all-ones mask.
Further, $\delta$ is either all-zeros or a machine word that holds a pseudo-random value with a Hamming weight of approximately half the machine word size.
When $a[i]$ and $b[i]$ are written back to memory, the Hamming distance of the old and new value is either zero or also corresponds to approximately half the machine word size.
The computation of $\delta$ and the store operations are repeated for all $n$ machine words.
Moreover, $\mathsf{CT\_SWAP}$ is executed for each coordinate of the elliptic curve point.
Therefore, the leakage amplification depends also on the used curve.
For 64-bit architectures and the \textit{secp521r1} curve, this amounts to the projective coordinates $X$, $Y$ and $Z$, each of which is stored in nine machine words and includes additional flags such as \textit{negative}, which is part of OpenSSL's \textit{bignum} representation.
For \textit{Ed25519}, only four machine words per coordinate are necessary.

After the publication of \cite{alam_nonce_at_once}, libgcrypt adopted a countermeasure, which we analyze in detail in \cref{sec:revisiting}.
Adopting a countermeasure for OpenSSL was discussed but never realized\footnote{\url{https://github.com/openssl/openssl/pull/14464}}\footnote{\url{https://github.com/openssl/openssl/pull/15702}}\footnote{\url{https://github.com/openssl/openssl/pull/16543}}.
Therefore, it is still vulnerable to the attack from~\cite{alam_nonce_at_once,nascimento_ecc_cmov_sca,nascimento_ecc_embedded_sca}.

\paragraph{Alternative implementations and attacks}
Alternative attacks on the double-and-always-add routine do not target the conditional swap, but the value of an intermediate coordinate by using online template attacks to recover whether an addition was executed on the input coordinate or not~\cite{batina_online_ta, dugardin_dismantling_ecc_ta, roelofs_online_ta_other_side}.
Some libraries such as BoringSSL and WolfSSL use a windowed multiplication with a precomputed table~\cite{bernstein_highspeed_signatures}.
Profiled and non-profiled attacks on such implementations were proposed in~\cite{weissbart_one_trace, jin_ecdsa_collision}.
The authors of \cite{medwed_ta_ecdsa} proposed a template attack that targets intermediate points on a curve to recover the first few bits of the nonce.
This could in theory work independently of the used multiplication algorithm.
Lastly, many libraries formerly used \ac{wnaf} and \ac{naf} double-and-add algorithms\footnote{e.g., OpenSSL until release \textit{1.1.1} (2018):\newline\url{https://openssl-library.org/news/openssl-1.1.1-notes/index.html}}, where the sequence of double-and-add operations leaks nonce bits.
This was exploited in \cite{belgarric_ecdsa_android,genkin_ecdsa_mobile_nonintrusive,genkin_ecdh_sca_pc}.
%
%

\paragraph{Target software}
The attacks from~\cite{alam_nonce_at_once,nascimento_ecc_cmov_sca,nascimento_ecc_embedded_sca} are the basis for our work.
In the following, we focus mainly on OpenSSL (v3.6.0) and the \textit{secp521r1} Weierstrass curve.
Additionally, we investigate libgcrypt (v1.9.4), which adopted an \ac{em} \ac{sca} countermeasure after publication of \cite{alam_nonce_at_once}, and \textit{Ed25519}.
It should be noted that while both OpenSSL and libgcrypt may choose to integrate countermeasures against \ac{em} \ac{sca}, they do not consider such attacks in their threat models.

\section{Revisiting Nonce@Once}
\label{sec:revisiting}
\silvan{}{Alternative title for section: Nonce@Once, Part I: Evaluation on a Baseline SoC}
\silvan{}{The term "Nonce@Once" was never defined in the paper"s main content. So it should be mentioned here}

In this section, we explore alternative methodologies for \ac{sca} of the conditional swap operation in \ac{ecdsa}~\cite{alam_nonce_at_once,nascimento_ecc_cmov_sca,nascimento_ecc_embedded_sca} and investigate countermeasures.
As baseline target we used a Raspberry Pi 4, which is much less complex than contemporary smartphones.
It is similar in complexity to the targets from~\cite{alam_nonce_at_once}, a ZTE ZFIVE and an Alcatel Ideal smartphone.

\paragraph{Measurement setup and target device}
We removed the metal lid to expose the die.
Images of the chip and die can be found in \cref{appendix:rpi4}.
The standard Raspbian \ac{os} was installed on the device.
It was connected to \ac{wifi} and we used \ac{ssh} to interact with the device.
Bluetooth was disabled.
We used a LeCroy WavePro 804HD oscilloscope and various \ac{em} probes from Langer EMV.


\subsection{Identification of activity-modulated signals}

The work in \cite{alam_nonce_at_once} showed that sampling in a narrow band around the device's clock frequency and processing the acquired signal with a sliding-median filter unveils sufficient information about the device's activity, i.e., arithmetic operations and the conditional swap operations were visible.
This was only partially reproducible with our setup.
In particular for a clock frequency of \SI{1.8}{\giga\hertz}, the \ac{cpu}'s activity is hardly traceable and noise dominates the signal.
\silvan{}{We should explicitly mention on our contribution here: We show a path on how to use the STFT to locate the scaled-down activity bands for modern SoCs.}

\begin{figure}
	\centering
	\includegraphics[width=\linewidth]{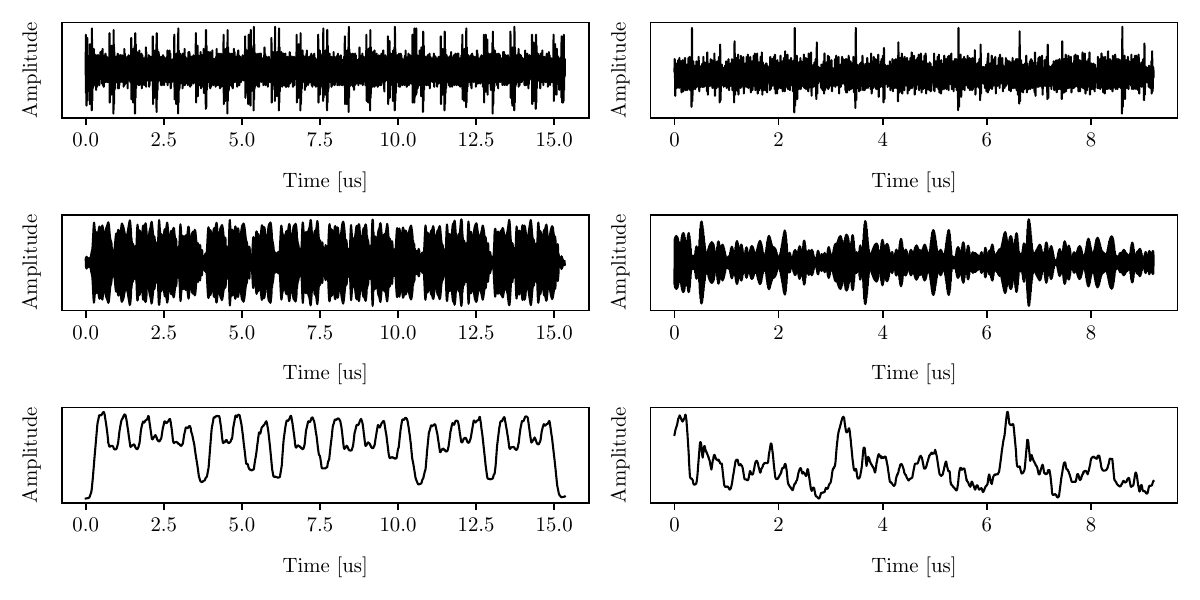}
	\caption{\ac{em} traces for a single Montgomery ladder step for OpenSSL + \textit{secp521r1} (left) and three iterations for the double-and-always-add routine of libgcrypt and \textit{Ed25519} (right). The top row shows the unfiltered signals. The traces in the middle row are bandpass-filtered for \SI{130}{\mega\hertz}. The traces in the bottom row are additionally absolute and sliding-median filtered. The Raspberry Pi 4's clock frequency was \SI{1.8}{\giga\hertz} for this investigation.}
	\label{fig:rpi4osslgcry}
\end{figure}

However, during our analyses, we noticed activity in frequency bands far below the \ac{cpu} frequency which -- nevertheless -- seems clearly related to the \ac{cpu} activity.
\cref{fig:rpi4osslgcry} shows the frequency-filtered signals for OpenSSL + \textit{secp521r1} and libgcrypt + \textit{Ed25519}.
In particular, the absolute and sliding-median filtering of bandpass-filtered signals is helpful to detect repeating patterns.
We used a window length of approximately \SI{2}{\micro\second} for the sliding-median filter.
For the \textit{secp521r1} curve in OpenSSL, the peaks of this signal form a \textit{5-2-1-2-3-1-3-3} pattern, which can be directly mapped to finite-field multiplication and square operations within the ladder step operation.
The ladder step implementation uses the differential addition-and-doubling formulas from~\cite{tetsuya_ec_mult} (\cref{appendix:montladderstep}).
For libgcrypt and \textit{Ed25519}, the pattern cannot be directly mapped to operations within the double-and-always-add routine and the relevant frequency band is less dominant in the spectrum.
We observed these effects for measurements at two locations.
We placed a Langer EMV RF-B 2-3 probe on the die, above the location of the four \acp{cpu} (\cref{fig:rpi4dieshot}), and a Langer EMV RF-U 2.5-2 probe on the center capacitor of the triplet marked with a blue rectangle in \cref{fig:rpi4chip}.
The signal on the capacitors looked more promising and the position is easier to reproduce, therefore we used it for all subsequent analyses.
The capacitors act as a low-pass filter, which attenuate spectral components with higher frequencies such as the \ac{cpu} frequency, but do not dampen the frequency bands we consider here.

The occurrence of distinct frequency bands for cryptographic operations was reported before, but its cause was never conclusively identified.
Previous works have shown that distinct spectral components of operations can be used to break \ac{rsa} and \ac{ecc} implementations with secret-dependent branches \cite{nakano_preprocessing_smartphones,genkin_ecdsa_mobile_nonintrusive,genkin_ecdh_sca_pc,belgarric_ecdsa_android}.
To trace the origin of these frequency bands, the authors of \cite{belgarric_ecdsa_android} recommended to explore the microarchitecture of processors.
\acp{cpu} with an ARM Cortex-A8 architecture have two \acp{alu}, only one of which contains a multiplier.
The alternating use of these \acp{alu} might explain the distinct pattern of finite-field multiply/square operations for OpenSSL and \textit{secp521r1}.
Other works -- related to \acp{sca} aiming to recover user activity~\cite{callan_fase, prvulovic_finding_fm_am, wang_finding_carriers}, covert channels~\cite{yilmaz_capacity_covert_channel_instructions}, and malware detection~\cite{vedros_code_to_em} -- observe that voltage regulator clocks and memory refreshes can also act as carriers for such activity-modulated signals.
However, both carriers and activity-modulated signals are hard to identify and clearly associate with a hardware component even for dedicated micro-benchmarks~\cite{wang_finding_carriers}.
Accordingly, it is difficult to assign the patterns in~\cref{fig:rpi4osslgcry} to a carrier frequency with a corresponding hardware component.
By artificially altering the \ac{cpu} frequency, we could verify that the frequency of the modulated signal had a constant factor of roughly $\nicefrac{1}{14}$ relative to the \ac{cpu} frequency, e.g., if we manually enforce a clock frequency of \SI{1.1}{\giga\hertz}, we observed the relevant activity around \SI{80}{\mega\hertz}.
This linear relationship suggests that a clock that scales linearly with the \ac{cpu} frequency or the \ac{cpu} frequency itself acts as carrier.

As long as the device is active (e.g., when a process like signature generation is triggered once in a while), it mostly uses a frequency of \SI{1.8}{\giga\hertz}.
The behavior described above would allow an attacker to easily adapt the attack to any \ac{cpu} frequency.

\subsection{Leakage assessment and exploitation}
\label{subsec:rpi4leakageassess}

The conditional swap operation~(\cref{alg:cswap}) occurs once before each step on the Montgomery ladder for OpenSSL or once after each iteration of the double-and-always-add routine in libgcrypt.
As discussed in \cref{subsec:vulnerability:libs}, the single-bit swap condition is amplified to multiple operations with distinct Hamming weights and distances.
OpenSSL still implements the swap operation as described in \cref{alg:cswap}, whereas libgcrypt adopted the countermeasure described in \cref{alg:cswapgcry}.

\begin{algorithm} 
	\caption{$\mathsf{CT\_SWAP}$ with \ac{em} \ac{sca} countermeasure as implemented in libgcrypt.}
\label{alg:cswapgcry}
\hspace*{\algorithmicindent} \textbf{Input:} Two arrays $a,b$ of size $n$ machine words, representing a coordinate of an elliptic curve point; Swap condition $cond \in \{0, 1\}$.\\
\hspace*{\algorithmicindent} \textbf{Output:}\\If $cond = 1$: $a \gets b$, $b \gets a$;\\If $cond = 0$: $a \gets a$, $b \gets b$.
\begin{algorithmic}[1]
\State $mask \gets 0 - cond$ \Comment{$mask$ is all-ones or zeros}
\State $\neg mask \gets \neg(0 - cond)$ \Comment{inverse $mask$}
\For{$i\gets0$ \textbf{to} $n-1$}
  \State $\delta_0 \gets (a[i] \; \& \; \neg mask) \; | \; (b[i] \; \& \; mask)$
  \State $\delta_1 \gets (a[i] \; \& \; mask) \; | \; (b[i] \; \& \; \neg mask)$
  \State $a[i] \gets \delta_0$
  \State $b[i] \gets \delta_1$
\EndFor
\end{algorithmic}
\end{algorithm}

\begin{algorithm}
	\caption{$\mathsf{CT\_SWAP}$ with \ac{em} \ac{sca} countermeasure as proposed in \cite{alam_nonce_at_once}.}
\label{alg:cswapwcountermeasure}
\hspace*{\algorithmicindent} \textbf{Input:} Two arrays $a,b$ of size $n$ machine words, representing a coordinate of an elliptic curve point; Swap condition $cond \in \{0, 1\}$; Random word $r$.\\
\hspace*{\algorithmicindent} \textbf{Output:}\\If $cond = 1$: $a \gets b$, $b \gets a$;\\If $cond = 0$: $a \gets a$, $b \gets b$.
\begin{algorithmic}[1]
\State $mask \gets 0 - cond$ \Comment{$mask$ is all-ones or zeros}
\For{$i\gets0$ \textbf{to} $n-1$}
  \State $\delta \gets (a[i] \oplus b[i]) \; \& \; mask$
  \State $\delta \gets \delta \oplus r$
  \State $a[i] \gets (a[i] \oplus \delta) \oplus r$
  \State $b[i] \gets (b[i] \oplus \delta) \oplus r$
\EndFor
\end{algorithmic}
\end{algorithm}

\begin{figure}
	\centering
	\includegraphics[width=\linewidth]{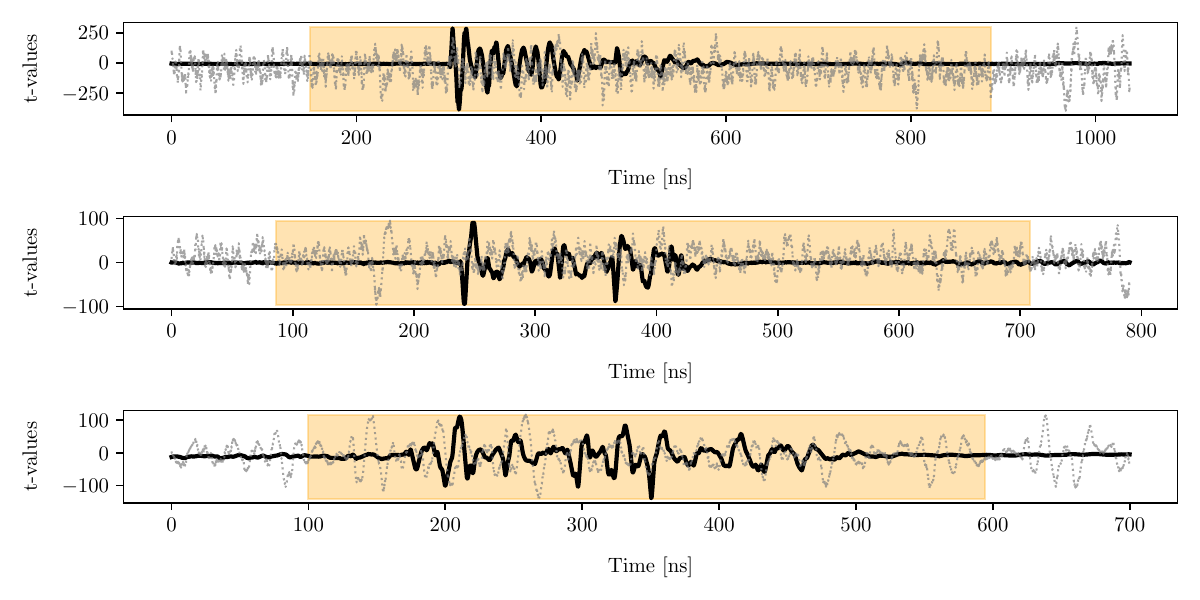}
	\caption{Leakage assessment results depicting t-values (solid, black line), a sample \ac{em} trace (dotted, grey line) and the trigger window (marked in orange). The top plot shows the results for OpenSSL +\textit{secp521r1}, the middle plot for libgcrypt + \textit{Ed25519} and the bottom plot for OpenSSL + \textit{secp128r1} with the countermeasure from~\cite{alam_nonce_at_once}.}
	\label{fig:rpi4leakage}
\end{figure}

The authors of \cite{alam_nonce_at_once} were able to identify visually recognizable differences in the amplitude of the \ac{em} signal depending on the swap condition.
Further, they relied on a $k$-means clustering algorithm with Euclidean distance metric to train a classifier capable of recovering the swap condition.

\paragraph{Leakage assessment}
We cannot directly classify traces visually with regard to their swap condition.
Therefore, we relied on established \ac{sca} methodologies such as Welch's t-test \cite{schneider_leakage_assessment}.
For this purpose, we recorded one million traces with a random, but known, swap condition.
We used the Raspberry Pi 4's \ac{gpio} pins to set a trigger signal at the beginning of the swap operation and reset it once it finishes.
To keep jitter minimal, we allowed the program to run two step operations of the Montgomery ladder before we set the trigger for the single swap operation.
If we isolate the swap operation, i.e., directly after the target program is started, the trigger pin is set and $\mathsf{CT\_SWAP}$ is executed, the standard deviation of the time the trigger signal is active is twice as high.
Otherwise, we followed the standard procedure for \ac{tvla}~\cite{schneider_leakage_assessment}, such as randomly selecting the swap condition for each trace to avoid microarchitectural effects.
A detailed study regarding best-practices for \ac{tvla} on complex devices with dynamic frequencies and cache memories could be beneficial for future work.
We used a sampling rate of \SI{10}{\giga\hertz} and did not apply any filtering or alignment.

\cref{fig:rpi4leakage} shows the results of our leakage assessment.
For all three of our experiments, we found a strong dependence of \ac{em} signals on the swap condition due to t-values far beyond the threshold of $|t| > 4.5$~\cite{schneider_leakage_assessment}.
This is obvious for OpenSSL + \textit{secp521r1}, as no countermeasure is in place.
The countermeasure libgcrypt adopted involves the inverse bit-mask $\neg mask$, which aims to eliminate certain parts of the leakage amplification, as both an all-ones and an all-zeros mask are used (\cref{alg:cswapgcry}).
However, since $mask$ and $\neg mask$ are processed at different times, the leakage amplification still exists, even if a higher temporal resolution is needed to measure it.
Further, the Hamming distance leakage of overwriting or not overwriting the arrays $a$ and $b$ in memory still exists.
The authors of~\cite{alam_nonce_at_once} proposed a different countermeasure described in \cref{alg:cswapwcountermeasure}.
It also eliminates part of the leakage amplification by introducing a random mask $r$.
However, $mask$ still amplifies the swap condition $cond$ and so does the Hamming distance leakage of conditionally updating the arrays $a$ and $b$.
We implemented this countermeasure in OpenSSL and applied the same constraints as~\cite{alam_nonce_at_once} to ensure that the compiler does not remove the countermeasure.
We selected \textit{secp128r1} as one of the curves with the least amount of leakage amplification.
According to the t-test, there is still a strong dependence of \ac{em} emanations on the swap condition.
Please note that the authors of \cite{alam_nonce_at_once} used a \ac{sdr} with a lower sampling rate, lower bandwidth, and less ideal measurement positions for the attack and the evaluation of their countermeasure.

\paragraph{Evaluating classifiers}
We evaluated several methods to classify traces by swap condition.
Neural networks, in particular, promise high accuracy and some immunity to temporal jitter~\cite{gohr_ches_challenge, picek_sok_deep_learning_sca}.
We trained classifiers on \num{800000}~traces of \num{1000}~sample points, which were chosen according to the t-test.
The remaining \num{200000} traces were used for validation.
For the evaluation of multiple classifiers, we used our measurements for OpenSSL + \textit{secp521r1}.
The highest accuracy we achieved is \SI{98.14}{\percent} for a custom \ac{cnn} architecture.
We derived the architecture by using a \ac{cnn} published as part of the \ac{ascad}~\cite{ascad} as basis for hyperparameter tuning.
More details on the \ac{cnn} architecture and the accuracy of other classifiers such as Gaussian templates can be found in~\cref{appendix:classifiers}.
For \textit{Ed25519} in libgcrypt, the classifier attained a maximum accuracy of \SI{98.03}{\percent}.

\subsection{Attack evaluation}
\label{subsec:rpi4attackeval}

For the attack, we first needed to identify all swap operations.
As discussed previously, we used activity-modulated signals to identify arithmetic operations.
We automated this process and, with a few manual corrections, could identify all swap operations.
For some traces we observed interruptions of the cryptographic operation, possibly related to actual interrupts, and interference signals, which do not interrupt the cryptographic operation, but overlay parts of the trace.
The former can be counteracted with manual correction.
The latter affects the probability with which the \ac{cnn} can correctly classify the swap, but typically only up to ten swap operations are affected.
For the classification of swap operations, we used the unfiltered signal, which was sampled at \SI{2.5}{\giga\hertz}.
Note that frequencies above \SI{1.25}{\giga\hertz} experience aliasing for this sampling rate.
This is inconsequential for our attack, because the activity‑modulated signals of interest lie in lower frequency bands and the classification accuracy is hardly impacted, as evidenced by the results below.

We analyzed four traces per library.
For OpenSSL and \textit{secp521r1}, we correctly obtained \num{496}, \num{508}, \num{509} and \num{517} of \num{521} bits.
For libgcrypt and \textit{Ed25519}, we recovered \num{207}, \num{210}, \num{226}, and \num{235} of \num{255} bits.
While the countermeasure complicates the attack, a single-trace attack remains feasible, albeit with increased computational effort.
Approaches to exploit multiple traces and signatures are discussed in~\cref{subsec:latticerecomb}.

\section{Nonce@Once on a modern smartphone}
\label{sec:contemporarysmartphones}
\silvan{}{Alternative title for section: Nonce@Once, Part II: Case Studies on Modern Smartphones}

In this section, we show two case studies, demonstrating the challenge when transferring the Nonce@Once attack to a modern smartphone target.
First, we target the Fairphone~4 running a native Linux \ac{os} that allows root access.
For the second case study, we use the Fairphone~4's stock Android and embed the cryptographic library into a custom Android app.
We focus only on OpenSSL and the \textit{secp521r1} curve.

\subsection{Case study 1: Fairphone 4 and native Linux}
\label{subsec:cs1}

The Fairphone 4's main \ac{soc} is the Snapdragon~750G~5G~\cite{snapdragon750}, which is also used in the Samsung Galaxy A52, Xiaomi Mi 10T, OnePlus Nord and other mid-range smartphones that were sold from 2021 and onward.
As such, it is representative of smartphones that are slowly fading from the market, but are in active use for a few more years.
As shown in~\cref{tab:sota}, it is one of the most complex devices ever investigated for \ac{em} \ac{sca} to date.
The goals of our first case study utilizing a Linux \ac{os} are to eliminate the complexity of the software stack as much as possible, to be able to apply constraints to scheduling, and to find simple solutions for triggering.
The Fairphone~4's open ecosystem which provides support for alternative operating systems and an open \ac{pcb} schematic\footnote{\url{https://www.fairphone.com/wp-content/uploads/2022/09/FP4_Information-for-repairers-and-recyclers.pdf}} facilitates our investigations.
However, we anticipate that our results translate to other devices based on the same or similar \acp{soc}.

\paragraph{Device preparation}
The Fairphone 4 is designed for repairability.
Therefore, it was easy to remove the main \ac{pcb} and disconnect unnecessary peripherals such as the camera.
The metal lids covering the \ac{soc} and its peripherals on the opposite side of the \ac{pcb} could be easily unclasped.
We removed the plastic package of the \ac{soc}'s \ac{ic} for some analyses, using an ULTRA TEC ASAP-1 CNC Mill.
We emphasize that an attacker could also use a cheaper stationary drill.
Except for the software, the laboratory setup was identical to the subsequent case study (\cref{fig:fp4setup}).
An \ac{ir} die-shot of the Snapdragon~750G~5G can be found in \cref{appendix:snapdragon750}.
The phone was connected via \ac{usb} and fully charged.
For this case study, we unlocked the phone's bootloader and installed postmarketOS\footnote{\url{https://postmarketos.org/}}.
This allows root access and limits the amount of background processes.
The device's \ac{wifi} interface was disabled and no \ac{sim} card was installed.

\paragraph{Triggering}
To facilitate our analyses, we explored various ways of firing a trigger signal from the phone.
We discovered that the \ac{i2c} communication between the \ac{soc} and the AW8695 \ac{ic}, which drives the phone's vibration motor, provided an efficient method for triggering.
The \ac{pcb} schematic helped to identify the \ac{i2c} bus's pull-up resistors and solder a thin wire to them.

\subsubsection{Investigation of \acp{cpu}}

\begin{figure}
	\centering
	\includegraphics[width=\linewidth]{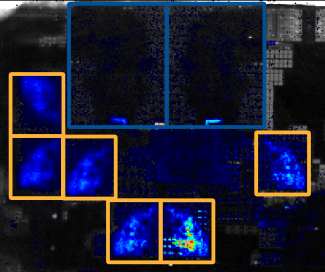}
	\caption{Excerpt of an \ac{ir} die-shot of the Snapdragon 750G 5G \ac{soc} overlaid with photon emission measurements for \ac{cpu} identification. In this case, a simple benchmark program is scheduled to one of the six A55 \acp{cpu} marked with orange rectangles. The two A77 \acp{cpu} are highlighted with blue rectangles.}
	\label{fig:fp4cpuid}
\end{figure}

The Fairphone 4's \ac{soc} incorporates two distinct \ac{cpu} variants.
The six Kryo 570 Silver / Cortex-A55 cores are optimized for energy efficiency, while the two Kryo 570 Gold / Cortex-A77 cores are optimized for performance.
Using a PHEMOS-X emission microscope, we recorded an \ac{ir} die-shot and the photon emission, when scheduling a program to a specific \ac{cpu}.
The results are shown in \cref{fig:fp4cpuid}.
The complete high resolution die-shot without the photon emission overlay can be found in \cref{appendix:snapdragon750}.

The attack must be tailored to the specific \ac{cpu} executing the cryptographic operation.
To investigate the electromagnetic profile of the \acp{cpu}, we placed a Langer EMV RF-B 3-2 probe over the respective area of the \ac{soc} and scheduled a scalar-by-point multiplication on the core.

The \acp{cpu} of the Snapdragon 750G 5G support a wide range of frequencies (\cref{appendix:snapdragon750}), with a maximal frequency of \SI{1.8}{\giga\hertz} for A55 and \SI{2.21}{\giga\hertz} for A77 \acp{cpu} and a minimal frequency of \SI{300}{\mega\hertz} for both.
When triggering the signature generation in intervals between one to three times per second, we observed that most of the time the A55 \acp{cpu} are clocked at \SI{768}{\mega\hertz} and the A77 \acp{cpu} are clocked at \SI{787}{\mega\hertz} while running the process.
We consider this to be a realistic approximation of the actual attack, in which only one signature is generated, or the \ac{gui} limits the rate in which signatures can be created.
We observed a shorter run time due to a higher core frequency in only about one in ten traces.
In contrast to the Raspberry Pi 4, the Snapdragon~750G~5G seems to scale its \ac{cpu} frequency more conservatively, most likely due to energy efficiency.

\begin{figure}
	\centering
	\includegraphics[width=\linewidth]{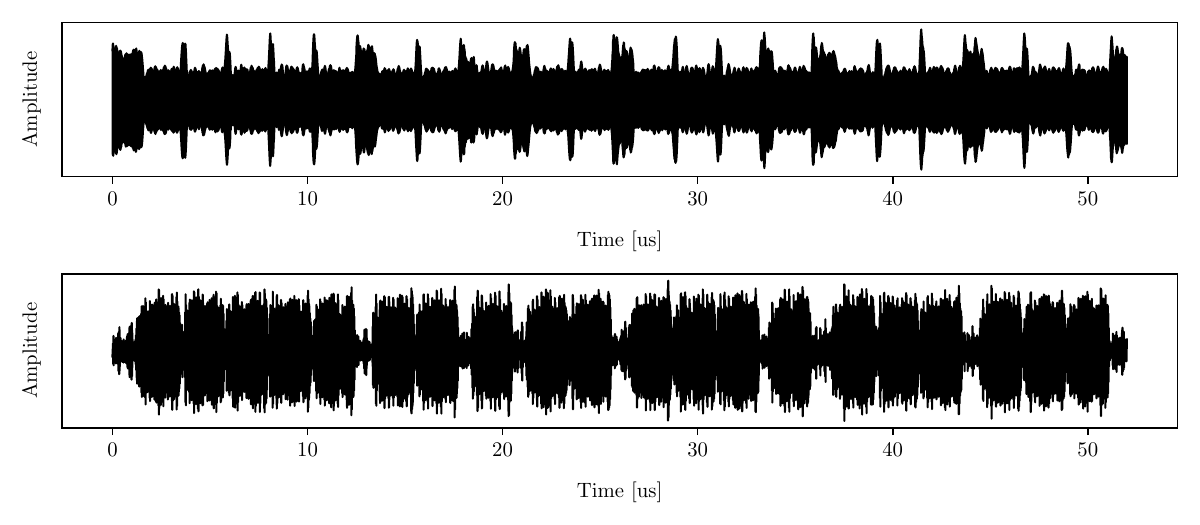}
	\caption{Single step on the Montgomery ladder on one of the Snapdragon 750G 5G's Cortex-A55 \acp{cpu} bandpass-filtered for \SI{768}{\mega\hertz} (top) and \SI{40}{\mega\hertz} (bottom).}
	\label{fig:fp4core0ladder}
\end{figure}

\begin{figure}
	\centering
	\includegraphics[width=\linewidth]{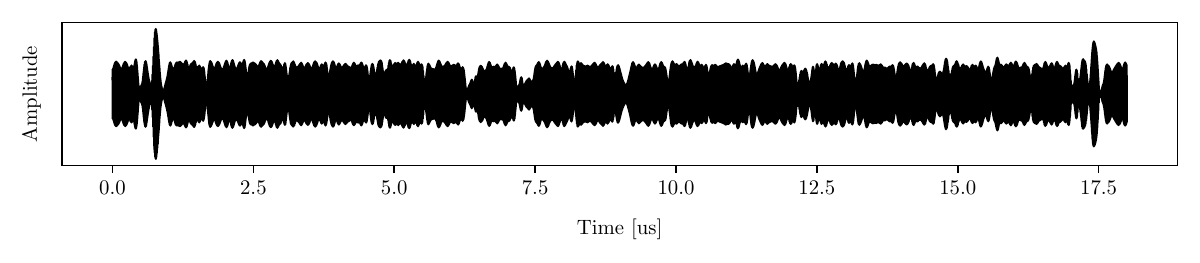}
	\caption{Single step on the Montgomery ladder on one of the Snapdragon 750G 5G's Cortex-A77 \acp{cpu} bandpass-filtered for \SI{787}{\mega\hertz}.}
	\label{fig:fp4core7ladder}
\end{figure}

\cref{fig:fp4core0ladder} and \cref{fig:fp4core7ladder} contain bandpass-filtered signals for a single ladder step on the Montgomery ladder if scheduled on an A55 or A77 \ac{cpu} with the core frequencies described above.
For the A55 core (\cref{fig:fp4core0ladder}), we observed a similar modulation effect as for the Raspberry Pi 4 at frequencies of $f_{mod_l} = \SI{41}{\mega\hertz}$ and $f_{mod_h} = \SI{82}{\mega\hertz}$.
Applying a \SI{41}{\mega\hertz} bandpass filter unveils the distinctive pattern of finite-field multiplication and square operations we also observed on the Raspberry Pi 4 (\cref{fig:rpi4osslgcry}).
When artificially forcing higher clock frequencies, we observed that $f_{mod_h} = 2 f_{mod_l}$ always holds and -- as observed on the Raspberry Pi -- the factors $\nicefrac{f_{mod_{l / h}}}{f_{cpu}}$ remain constant.
Since we were measuring on the die and not -- as for the Raspberry Pi 4 -- on a capacitor, which attenuates higher frequencies, we could also extract information from a signal that has been filtered for the \ac{cpu} frequency of \SI{768}{\mega\hertz}.
In this case, the amplitude is increased for the time intervals between finite-field multiplications and square operations.

\begin{figure}
	\centering
	\includegraphics[width=\linewidth]{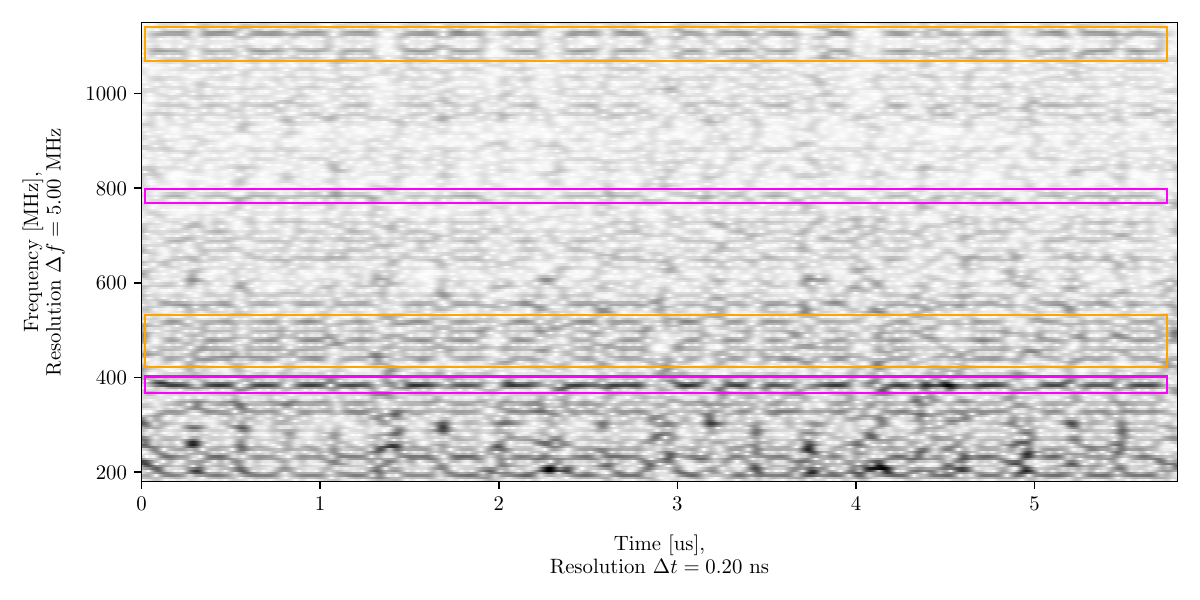}
	\caption{\ac{stft} for a single step on the Montgomery ladder on one of the Snapdragon 750G 5G's Cortex-A77 \acp{cpu} with a frequency of \SI{2.21}{\giga\hertz}. The distinct pattern of repeating blocks in a \textit{5-2-1-2-3-1-3-3} structure can be observed on multiple frequency bands.}
	\label{fig:fp4core7stft2210mhz}
\end{figure}

For the A77 \ac{cpu} (\cref{fig:fp4core7ladder}) and a \ac{cpu} frequency of \SI{787}{\mega\hertz}, we get the best signal by applying a bandpass filter around the \ac{cpu} frequency.
\cref{fig:fp4core7stft2210mhz} shows the \ac{stft} result when forcing the \ac{cpu} frequency to the maximum of \SI{2.21}{\giga\hertz}.
In this case, filtering for the \ac{cpu} frequency leads to noisy traces, from which operations are not directly apparent.
However, we detected two kinds of useful activity-modulated signals.
Symmetric to $\nicefrac{1}{2} f_{cpu} =$\SI{1.105}{\giga\hertz} and $\nicefrac{1}{4} f_{cpu} =$\SI{552.5}{\mega\hertz}, we observed the modulated activity in two, respectively four bands (marked in orange in~\cref{fig:fp4core7stft2210mhz}).
This seems to indicate that the \ac{cpu} frequency serves as the carrier and that the cryptographic activity is modulated onto various harmonics, which scatter for higher-order harmonics.
Additionally, we observed activity-modulated signals at \SI{390}{\mega\hertz} and \SI{790}{\mega\hertz} (marked in magenta in~\cref{fig:fp4core7stft2210mhz}).
By varying the \ac{cpu}'s frequency, we confirmed that the factors $\nicefrac{f_{mod}}{f_{cpu}}$ were also constant here.

\silvan{}{To Figure 6 caption: maybe label the frequency-bands?}

We were thus able to confirm that helpful activity-modulated signals exist at frequencies well below the \ac{cpu} frequency for both \ac{cpu} variants present in the Snapdragon 750G 5G.
This is particularly important for high \ac{cpu} frequencies, where we observe significant attenuation by the package when measuring with and without the package.
Our results confirm that the carrier is most likely the \ac{cpu}'s clock frequency.
For some spectral components (e.g., the ones marked in orange in \cref{fig:fp4core7stft2210mhz}), the factor $\nicefrac{f_{mod}}{f_{cpu}}$ is easy to explain, whereas for others (e.g., the ones marked in magenta in \cref{fig:fp4core7stft2210mhz}), we can only assume that the cause lies in the microarchitecture of the \ac{cpu} combined with periodicity in the running software.

\subsubsection{Results for Nonce@Once attack}

To reduce the complexity of our first case study, we constrained the scheduling via \textit{taskset} such that a fixed A55 \ac{cpu} was used for the scalar-by-point multiplication.
This simplified the attack significantly, as we could place the probe accordingly.
The increased complexity of a mobile platform and the more advanced semiconductor node are the most impactful challenges in this case study.
We used the device for which we removed the package.
As in \cref{subsec:rpi4attackeval}, we used a sampling rate of \SI{2.5}{\giga\hertz}.

Although our \ac{i2c} triggering mechanism works reliably, its repetition rate is several orders of magnitude lower than that of a conventional \ac{gpio} trigger and it introduces more jitter.
To acquire a sufficient amount of reliable profiling data, we slightly modified the Montgomery ladder code we ran on the device during the profiling phase to extract as many swap operations as possible per trace and label them correctly.
In short, we used only nonces that correspond to always or never swapping and random base points for the scalar-by-point multiplication.
We acquired two million training traces in less than \SI{24}{\hour}.
The trained \ac{cnn} achieved a final accuracy of \SI{99.8}{\percent} during validation.

For the attack phase, we ran the unmodified OpenSSL code.
We used the artificial trigger signal for convenience.
Our results show that an amplitude-based trigger on a bandpass-filtered signal should also be sufficient.
The alignment of swap operations worked similar to the one described in \cref{subsec:rpi4attackeval}, using the activity-modulated signal around \SI{41}{\mega\hertz} and similar absolute, sliding-median filters.
Overall, we identified slightly more interruptions and interference signals than on the Raspberry Pi 4.
We evaluated the attack on two traces, where we successfully recovered \num{505} and \num{517} bits of the respective nonces.

Even though the initial investigation required more effort, the attack in this scenario is ultimately just as easy to carry out as on the Raspberry Pi 4.
\ac{cpu} pinning and a favorable measurement position on the die make the attack easier.
The A55 core's comparatively low operating frequency also helps.
However, this is not due to artificial constraints, but reflects normal dynamic frequency scaling, where the \acp{cpu} increase frequency only under higher load.

\subsection{Case study 2: Fairphone 4 and Android}
\label{subsec:cs2}

In this case study, we examine how the full Android software stack affects the feasibility of \ac{sca}.
Our target is a simple, custom Android app that includes native OpenSSL code.
Please note that in general, developers are recommended to use the Android Keystore.
We discussed our motivation for investigating libraries such as OpenSSL in \cref{subsec:vulnerability:libs}.

\begin{figure}[h!]
	\centering
	\includegraphics[width=\linewidth]{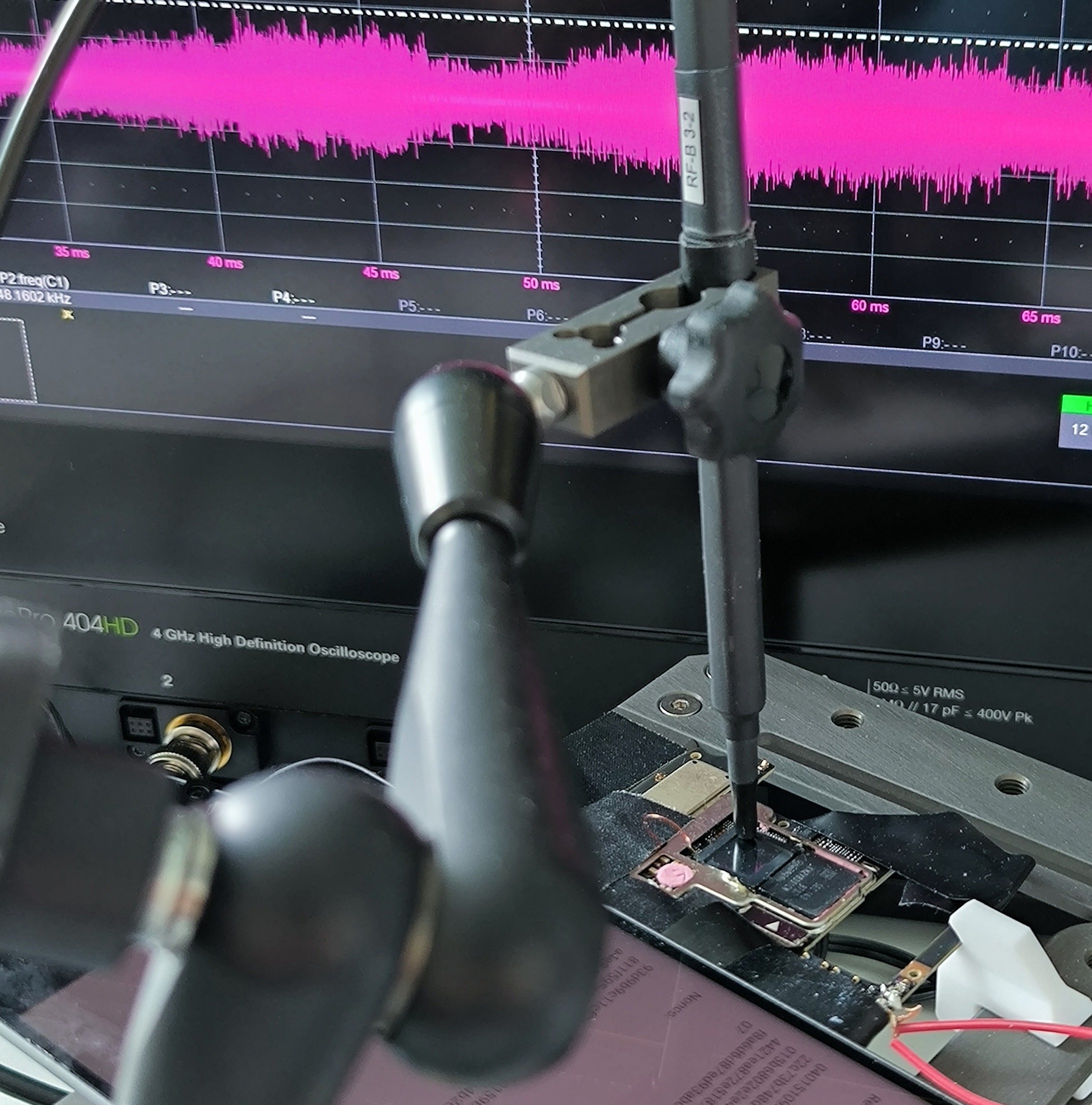}
	\caption{Measurement setup for \ac{em} \ac{sca} of the Fairphone~4 app target.}
	\label{fig:fp4setup}
\end{figure}

\paragraph{Setup}
The setup is shown in \cref{fig:fp4setup}.
The hardware was the same as in the previous case study.
We flipped the phone's display to place a probe on the \ac{soc}'s die.
For convenience, we again used an artificial trigger signal, the necessity of which will be discussed later.
We could not use the \ac{i2c} bus from the first case study, since Android uses it independently for other functions.
Instead, we relied on the \ac{pwm} signal that drives the phone's vibration motor, as it is simple to trigger a vibration from our target app.
The phone was running an unmodified Android \num{13}.
\ac{wifi} was not connected but enabled, no \ac{sim} card was installed.
The phone is connected to a host \ac{pc} via \ac{usb} and charging.
We used \ac{adb}'s \textit{uiautomator} to interact with the target app.

\subsubsection{Scheduling investigation}

Without the scheduling constraints of the previous case study, the first challenge is to understand where the cryptographic operation is scheduled.
First, we analyzed the photon emission, which we also used to locate individual \acp{cpu}, in order to obtain information about the scheduling behavior.
During this analysis, the phone was fully charged and connected to power.
Our main findings are that the device utilizes one of the A77 \acp{cpu} for Android's background tasks.
Individual native processes with computation times below \SI{1}{\second} are executed on the same core.
It is unclear why the system utilizes a performance core instead of an energy-optimized \ac{cpu} for this minimal load, but the full-charge status and connection to a power supply might be the cause for this.
For longer processes, we observed that an additional core, mostly one of the A55 \acp{cpu}, was used.
Further, we investigated the impact of Android apps, the load introduced by handling touch events, and the scheduling of our target app with native OpenSSL code.
We observed that an Android app, even if idle, caused increased activity on three A55 and one A77 \acp{cpu}.
Handling touch events further increased the activity on A55 \acp{cpu}, however we cannot make any statements about where cryptographic operations triggered by a touch event are scheduled.

We conducted \ac{em} measurements to gain further insights.
Although the \SI{2}{\milli\meter} resolution of the Langer EMV RF-B 3-2 probe is too fine to cover the complete area occupied by \acp{cpu} (\cref{fig:fp4cpuid}), we found a position where we can measure the \ac{em} emanations for all eight \acp{cpu}.
The use of Android and an app significantly increases the noise in the traces.
\cref{fig:fp4appladdertraces} shows that the operation is impossible to detect without a filter and is also potentially overlaid by noise signals.
However, we could recognize the characteristic activity-modulated signals we established in~\cref{subsec:cs1}.
We automated the detection of the relevant patterns in the characteristic frequency ranges and applied it to \num{100} traces.
For \num{57}, we could identify the Montgomery ladder on an A77 \ac{cpu} clocked at \SI{787}{\mega\hertz}.
For the remainder of traces, the Montgomery ladder was either scheduled on an A55 \ac{cpu}, the A77 core was clocked at a different frequency, or the trace was too noisy for automatic detection.

This makes the A77 \ac{cpu} with a frequency of \SI{787}{\mega\hertz} the most likely scheduling scenario.
At least, this holds for scenarios similar to our laboratory setup, i.e., the phone is fully charged and connected to power.
We also suspect that the app design has an impact, e.g., if it causes a higher load due to other activities, the clock frequency might be higher.

\subsubsection{Results for Nonce@Once attack}
We decided to tailor the attack to the A77 \ac{cpu} and a core frequency of \SI{787}{\mega\hertz} in this case study.
This means that we ignored traces where the A77 \ac{cpu} used a different frequency or the cryptographic operation was scheduled on an A55 \ac{cpu}.
We identified such traces from their characteristic spectral components and the duration of the operation.
As we showed above, scheduling on an A77 core with a frequency of \SI{787}{\mega\hertz} is the most likely scenario for our setup and target app.
For an actual attack, an adversary should either prepare classifiers for all \ac{cpu} variants and their most likely frequencies or be certain that enough traces can be recorded such that at least for one trace, the cryptographic operation is scheduled on the desired \ac{cpu} with the desired frequency.
However, the number of necessary traces is not only relevant for different scheduling scenarios, but also depends on the number of bits that can be recovered per trace.
We discuss this in more detail below.

\paragraph{Training trace acquisition and classifier training}
The easiest method for acquiring training traces would be to again use postmarketOS.
This would allow us to enforce the scheduling on the desired \ac{cpu}, enforce the frequency, and reduce the measurement time significantly, as the \ac{i2c}-based trigger could be used.
However, this might not be possible for every phone, therefore we explored the effort needed to acquire enough training traces on a phone running Android and our target app.
The advantage of this is that effects such as random use of one of the two A77 \acp{cpu} and increased noise due to more background activity are covered by the training data in a natural way.
The usage of the vibration motor as trigger signal limits the amount of measurements to a single trace per second.
Further, as discussed above, we discarded traces with non-matching scheduling conditions.
For training trace acquisition, we re-used the modifications to the Montgomery ladder described in~\cref{subsec:cs1} and obtained around two million training traces in five days of measurement time.
As in all previous attacks, we used a sampling rate of \SI{2.5}{\giga\hertz}.
The trained \ac{cnn} achieved a validation accuracy of \SI{92}{\percent}.
This is significantly lower than in the previous case study.
Possible explanations for this are increased noise, the complexity of the A77 \ac{cpu}, and the variable scheduling between the two A77 \acp{cpu}.

\begin{figure}
	\centering
	\includegraphics[width=\linewidth]{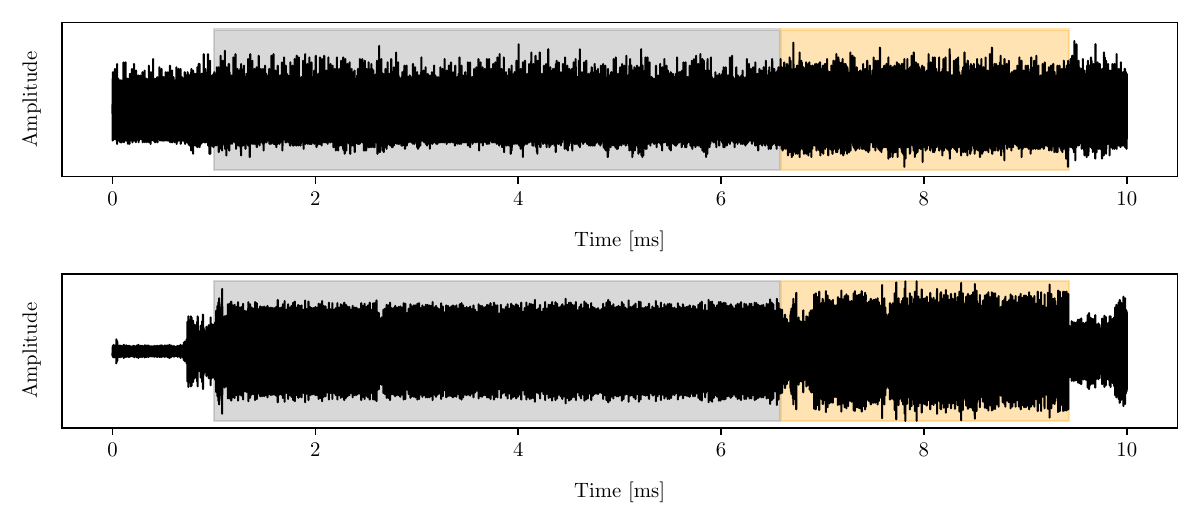}
	\caption{Trace of the scalar-by-point multiplication within an Android app without (top) and with (bottom) bandpass-filter for \SI{787}{\mega\hertz}. For the sub-trace highlighted in grey, swap alignment and recovery works with a high probability. An interference signal of unknown origin prevents this for the sub-trace highlighted in orange.}
	\label{fig:fp4appladdertraces}
\end{figure}

\paragraph{Attack phase}
For the attack phase, we recorded \num{20} traces.
We used our artificial vibration trigger.
Triggering on a bandpass-filtered \ac{em} signal should be possible; alternatively, pattern-based triggering could be explored.
Of the \num{20} traces, \num{11} seem to be applicable to our classifier, i.e., an A77 \ac{cpu} clocked at \SI{787}{\mega\hertz} is used.
Most of these traces are affected by intermittent interference signals as shown in \cref{fig:fp4appladdertraces}.
This increases the manual effort required to align swap operations.
We analyzed two traces for which we recovered \num{416} and \num{387} of the \num{521} bits correctly.
Both traces include interference signals as depicted in \cref{fig:fp4appladdertraces}.
We cannot definitively identify the cause of the interference.
It could be related to a frequency scaling or activity from another \ac{cpu}.
Within the superimposed part of the trace, the success rate is equivalent to randomly guessing the bit.
For the remainder of the traces, where swap operations are aligned with a high confidence, the \ac{cnn} almost achieves the same accuracy as during validation.
With $521-416 = 105$ or more incorrect/missing bits, obtaining the complete nonce in order to derive the secret key from the signature is no longer trivial.


\subsubsection{Discussion of key recovery methods}
\label{subsec:latticerecomb}
We assume that \num{416} and \num{387} bits per trace are not tight upper and lower bounds, respectively, on the number of bits that can be recovered.
Since most of the \num{11} traces show similar interference patterns, we used these values as the basis for analyzing key recovery methods.
With the given classifier and its validation accuracy, we expect approximately $521 \cdot 0.08 \approx 42$ bits to be erroneous, even if a trace is not overlaid by any interference signal.
First, we analyze to what extent the \ac{cnn}'s score/probability values for each guess can be used as an indication of potentially wrong nonce bits.
Knowing the indices of these bits is the difference between guessing \num{42} bits or trying $\binom{521}{42} \approx 2^{206.7}$ combinations for an interference-free trace.

\begin{figure}
        \centering
        \includegraphics[width=\linewidth]{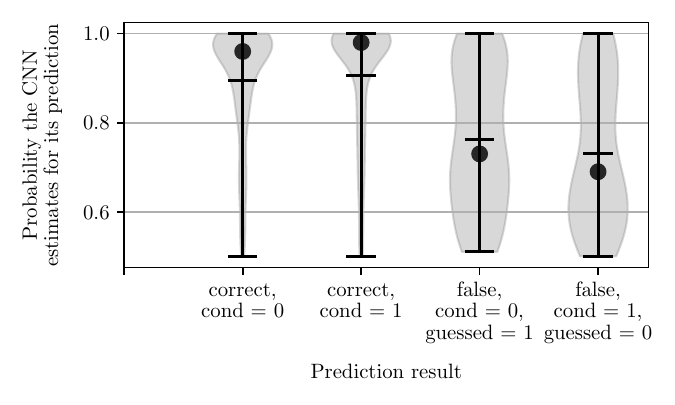}
        \caption{Violin plot \cite{hintze_violin} of the distribution of probabilities the \ac{cnn} estimates for its predictions. The vertical line highlights the distribution's mean, the circle marks the distribution's median.}
        \label{fig:fp4guesseval}
\end{figure}

\cref{fig:fp4guesseval} shows that the \ac{cnn}'s average probability value is around $0.9$ for correct and $0.75$ for incorrect guesses.
Therefore, many incorrect bits can be identified from the \ac{cnn}'s confidence, but the standard deviation of the probabilities for incorrect guesses shows that not all incorrect bits can be identified reliably.
This increases the brute-force effort beyond \num{42} bits.
Thus, even a completely interference-free trace is not guaranteed to enable a trivial brute-force recovery of the secret key.

If we assume that -- due to interference -- at most \num{416} bits per trace can be recovered, it is obvious that brute-force recovery is not feasible.
In this case, multiple traces, where each trace corresponds to a dedicated nonce - signature pair, can be utilized for the attack.
From the partial information about the nonce $k$ for each respective trace and the signature equation (\cref{subsec:ecdsa}), a \ac{hnp}~\cite{boneh_hnp} can be formulated, where the secret key $d_A$ corresponds to the hidden number.
This \ac{hnp} instance can be solved either by Fourier analysis~\cite{bleichenbacher, aranha_ladder_leak, demulder_bleichenbacher} or lattice-based recombination~\cite{albrecht_lattice_barrier,sun_guessing_bits,xu_lattice_sieving,gao_lattice_sieving}.
In the following, we conduct a short study based on recent work on lattice-based recombination~\cite{gao_lattice_sieving}.
In general, approaches based on Fourier analysis are believed to enable key recovery in scenarios where the number of leaking bits is too low for lattice-based recombination.
However, recent work claimed to close this gap for lattice-based recombination~\cite{gao_lattice_sieving}.
Further, lattice-based recombination usually requires significantly fewer signatures, which -- considering our threat models -- is an important metric.

Lattice-based recombination transforms the \ac{hnp} into a \ac{usvp}.
The way the lattice is constructed and then solved for the target vector was discussed in multiple works \cite{albrecht_lattice_barrier,sun_guessing_bits,xu_lattice_sieving,gao_lattice_sieving}.
The latest results \cite{gao_lattice_sieving} improved the lattice construction and algorithms originally proposed in \cite{albrecht_lattice_barrier} and covered a wide range of possible attack scenarios.

The challenge in applying their work to our scenario lies in the handling of errors.
The error rate in \cite{gao_lattice_sieving} was defined for each equation related to the \ac{hnp} and used to construct the lattice.
This means the probability of error $p_e$ is related to the probability that at least one of the $l$ recovered nonce bits for one nonce - signature pair is incorrect.
The success rate we used so far in this paper is per-bit, i.e., $p_b$ is the probability that a single bit is correct.
Since we recover each bit individually, $p_{e} = 1-(p_{b})^l$.
It is obvious that, even for significantly higher $p_b$ than we currently achieve, for large $l$, $p_e$ will always be relatively high, e.g., for $p_b = 0.99$ and $l = 100$, $p_e = 0.63$.

An additional constraint is that solvers from literature assume the $l$ bits to be in a consecutive chunk starting from the nonce's most- or least-significant bit.

\begin{figure}
        \centering
        \includegraphics[width=\linewidth]{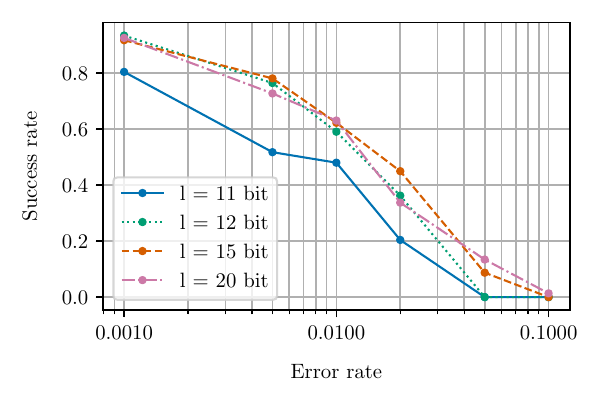}
        \caption{Success rate over error rate $p_e$ for key recovery by lattice recombination using the solver from \cite{gao_lattice_sieving}. We used \num{50} nonce - signature pairs (with $x = 1$) where $l$ bits of the nonce are known. The results are for the \textit{secp521r1} curve. For each parameter set, the success rate was calculated for \num{100} trials.}
        \label{fig:fp4latticesuccessrate}
\end{figure}

We used the solver from \cite{gao_lattice_sieving} to evaluate which error rates can be tolerated for $l \leq 20$, using \num{50} nonce - signature pairs\footnote{We use $x=1$, which makes the lattice equivalent to the one proposed in~\cite{albrecht_lattice_barrier}. While greater values for $x$ could be beneficial for our scenario, they massively increase the number of required signatures to a point where our specific attack would clearly be infeasible.}.
The results are shown in \cref{fig:fp4latticesuccessrate}.
We can see that, for a good chance at key recovery, $p_e \leq 0.01$ is required.
For $l=20$ this means $p_b > 0.9995$, and for $l=11$, $p_b > 0.999$ is needed.

From the data used for \cref{fig:fp4guesseval}, we evaluated the feasibility of pruning, i.e., not considering bits for which the \ac{cnn}'s confidence is below a certain threshold.
For example, if we only consider bits for which the \ac{cnn}'s probability is $0.99$ or higher, we get $p_b = 0.946$.
In this case, we must abandon \SI{65.7}{\percent} of bits during pruning.

This shows that the success rates we achieved in our laboratory setup are not sufficient for key recovery with the parameters used for \cref{fig:fp4latticesuccessrate}.
For $l \leq 10$, this might change, but such parameter choices make the computation highly time-consuming.
For reference, we cite a result from~\cite{xu_lattice_sieving}, who estimate that for a 521-bit modulus and $p_e = 0$, at least 66 and 132 signatures are needed for 8- and 4-bit leakage, respectively.

We conclude that a solver suitable to our scenario is currently missing in literature, i.e., a solver that can tolerate higher error rates and is more flexible with regard to the position of the leaking bits.
Our current experimental results are not directly applicable to existing solvers, but serve as a benchmark for the accuracy a measurement setup must achieve.

\section{Impact on smartphone security}
\label{sec:impact}

In this section, we broaden our focus and discuss the implications of our results for smartphone security. 

\subsection{Limitations}

Our evaluation has several practical limitations that affect the generality of the results.

\paragraph{Invasiveness.}
The attack relies on invasive measurements requiring physical modification of the device. 
In \cite{alam_nonce_at_once}, the Nonce@Once attack was demonstrated on older smartphones, where an \ac{em} probe placed on the device's casing was sufficient.
This would allow for scenarios in which the victim remains in possession of the device and does not notice the attack (\cref{subsec:scalocked}).
Therefore, we examined less invasive probe placements for the Fairphone~4.
For both measurements on the \ac{ic} and decoupling capacitors, we observed that the metal lids on top of them block the signal such that spectral components we use for alignment are almost completely attenuated.
We did not observe changes in the amplitude of measurements when starting a process.
The touchscreen above the \ac{ic} and the casing on top of the capacitors eliminated the spectral components completely.
While more advanced probes or signal processing techniques may improve this, demonstrating a fully non-invasive attack remains an open problem.

Apart from the probe position it is also interesting to note that the authors of \cite{alam_nonce_at_once} used a relatively inexpensive \ac{sdr} with a bandwidth of only \SI{40}{\mega\hertz}.
In comparison, we use more sophisticated laboratory equipment.
However, compared to the probe placement, this has less influence on the threat posed by \ac{em} \ac{sca} to smartphones.
At this point, we also emphasize that we used an artificial trigger for all our analyses.
This should be relatively easy to replace with pattern-based triggering on bandpass-filtered signals.

\paragraph{Transferability.}
Our experiments use a template attack with profiling and attack traces collected on the same device. 
Cross-device transferability would likely require more robust feature extraction and training across multiple devices, due to manufacturing differences and inaccurate probe placements.
Wrt. transferability to other elliptic curves, our analysis in \cref{sec:revisiting} indicated that the attack should also work for curves with lower leakage amplification.

\paragraph{Generality.}
We evaluated a single mid-range smartphone platform. 
While the underlying leakage mechanisms are expected to persist across similar architectures, modern smartphones increasingly employ heterogeneous multi-core designs, higher clock frequencies, and smaller technology nodes.
This complicates drawing general conclusions about whether our findings are applicable to other devices with different chipsets.

Another potentially influential factor is \ac{pop} architectures, in which memory is mounted on top of the \ac{ic}.
These are becoming increasingly common in smartphones and could make attacks much more difficult, as local measurements on the \ac{ic} are no longer possible without additional physical preparation~\cite{lisovets_iphone_bootloader, vasselle_bootloader, haas_apple_vs_ema}.

Regarding the transferability to a concrete app implementation, we want to emphasize that a multitude of operational conditions play a critical role.
Depending on the app architecture, the smartphone’s battery level, and background activity -- including that of the app itself and the \ac{os} -- clock frequency and scheduling may vary more significantly.
For newer smartphones, which sometimes feature up to three different \ac{cpu} variants and support a broader range of frequencies, even exceeding \SI{3}{\giga\hertz}, this results in a vast evaluation space.
However, for an attacker with high attack potential as it is for example assumed in~\cite{eucc_2025}, this additional effort is expected to have no significant impact on the attack's assessment.
Our investigations in \cref{subsec:cs1} showed that activity-modulated signals in lower frequency bands can help implement the attack for higher clock frequencies such as \SI{2.21}{\giga\hertz}.

Finally, it should be noted that Android applications should use the Android Keystore, which -- for most devices -- uses a proprietary library within a \ac{tee} or StrongBox.
In this case, the feasibility of attacks depends heavily on the implementation within this library and whether it includes countermeasures.

\paragraph{Complete key recovery}
Our analysis in \cref{subsec:latticerecomb} highlighted that a complete key recovery is not trivial for measurements of an Android app.
The investigation of solvers that can efficiently handle noisy leakage and the improvement of this leakage remain open challenges.

\subsection{Threat assessment}

The applicability of our investigated attacks strongly depends on the threat model. 
For the scenario of an unlocked smartphone under adversarial control (\cref{subsec:scaunlocked}), our case studies directly apply.
In this setting, an attacker can trigger cryptographic operations and potentially recover \ac{ecdsa} private keys, enabling unauthorized signing.
As discussed in \cref{subsec:scaunlocked}, stolen, but locked devices may fall into this category, if the lock mechanism can be bypassed.
If it cannot be bypassed, the attack surface is reduced to cryptographic operations which the locked phone might conduct automatically (\cref{subsec:scalocked}).
This highlights that strong locking mechanisms, which are also used to gate access to cryptographic operations, are an effective first line of defense.

For scenarios in which the device remains in possession of the user (\cref{subsec:scalocked}), the attack is currently not practical. 
Our measurements rely on invasive probe placements that require disassembling the device and removing shielding components.
This excludes realistic attack scenarios in which the adversary has no prolonged physical access to the device.
Here we assume that the attacker is not the owner of the device (\cref{subsec:scaunlocked}).

Overall, our results demonstrate that hardware side-channel attacks are difficult to mitigate purely in software. 
For critical identity solutions such as the \ac{eudi} wallet, short-term mitigation includes algorithmic countermeasures and protocol-level protections such as requiring a second factor (e.g., an eID card). 

\subsection{Countermeasures}
In the course of our analysis of the attack methodology, we found that the countermeasure libgcrypt adopted does not provide sufficient protection -- at least in the context of our laboratory setup and the relatively simple Raspberry Pi 4 platform.
Our leakage assessment results suggest that this also applies to the countermeasure described in~\cite{alam_nonce_at_once} (\cref{subsec:rpi4leakageassess}).
Consequently, the question of effective mitigation remains unresolved.
As described in \cref{subsec:vulnerability:libs}, widespread countermeasures such as coordinate re-randomization, point blinding, and scalar blinding are not or only partially effective against the attack.
Swapping pointers instead of the actual coordinates could reduce leakage amplification, but would introduce secret-dependent memory access patterns~\cite{alam_nonce_at_once}.
The authors of \cite{batina_sca_secure_ecc_sok} propose to combine scalar blinding, address randomization~\cite{itoh_countermeasure_address_dpa, heyszl_localem}, and projective coordinate re-randomization~\cite{coron_resistance_dpa_ecc}.
Scalar blinding implies that an adversary must recover all swap operations with a high accuracy, since only then it is possible to recover the actual scalar from the blinded one. 
Further, address randomization implies that two swap conditions must be recovered successfully per bit.
The leakage amplification for both swap conditions can be reduced, if projective coordinate re-randomization, and the countermeasure from~\cite{alam_nonce_at_once} are combined carefully.
That is, each coordinate must be completely re-randomized before it is written back, ensuring that values in memory are always updated.
The practical validation of countermeasures is left open as future work.


\section{Conclusion}
\label{sec:conclusion}

In this work, we conduct an analysis of the practicality of \ac{em} \ac{sca} on modern smartphones, utilizing the Fairphone~4's Snapdragon~750G~5G \ac{soc} as an exemplary target platform.
Our main contributions include the adaptation and execution of conditional-swap \ac{sca} on \ac{ecc}~\cite{alam_nonce_at_once,nascimento_ecc_cmov_sca,nascimento_ecc_embedded_sca} against recent hardware and software stacks, demonstrating that such attacks remain a tangible threat to contemporary smartphones.
Our case studies demonstrate that challenges such as heterogeneous \ac{cpu} clusters, dynamic frequency scaling, scheduling, and parallel activities of the device can be overcome.
We further show that the countermeasures proposed by \cite{alam_nonce_at_once} and implemented in libgcrypt are insufficient in practice.
Notably, this is the first work to systematically investigate a device as complex as the Fairphone~4 for \ac{em} \ac{sca}, analyzing full app contexts rather than isolated cryptographic libraries.
Beyond our practical investigations, we provide an overview of how cryptography is implemented on Android smartphones and where potential vulnerabilities lie.
We define two realistic threat models that guided our evaluation and allowed us to assess the real-world impact of our attack approaches.

Looking ahead, our findings make clear that understanding the full attack surface of smartphones is only just beginning.
Many open questions remain, in particular regarding the feasibility of attacks that do not require the smartphone to be disassembled and other limitations listed in~\cref{sec:impact}.
Our work serves as an important step towards a deeper understanding of real-world risks, and we hope it motivates the community to further explore side-channel security on complex mobile platforms.

Finally, we see our work in the context of critical identity solutions akin to the \ac{eudi} wallet.
Our results show clearly that such applications must be implemented with great care, and that in the long run, all smartphones need to be equipped with certified, secure hardware in which users can place their trust.

\bibliographystyle{IEEEtran}
\bibliography{references}

@inproceedings{alam_nonce_at_once,
  author       = {Monjur Alam and Baki Berkay Yilmaz and Frank Werner and Niels Samwel and Alenka G. Zajic and Daniel Genkin and Yuval Yarom and Milos Prvulovic},
  title        = {{Nonce@Once}: {A} Single-Trace {EM} Side Channel Attack on Several Constant-Time Elliptic Curve Implementations in Mobile Platforms},
  booktitle    = {2021 {IEEE} European Symposium on Security and Privacy (EuroS{\&}P)},
  pages        = {507--522},
  publisher    = {IEEE},
  year         = {2021},
  url          = {https://doi.org/10.1109/EuroSP51992.2021.00041},
  doi          = {10.1109/EUROSP51992.2021.00041}
}

@misc{eidas,
  author = {{Council of the European Union}},
  title  = {{Regulation (EU) 2024/1183 of the European Parliament and of the Council of 11 April 2024 amending Regulation (EU) No 910/2014 as regards establishing the European Digital Identity Framework}},
  year   = {2024},
  howpublished = {Online},
  url    = {http://data.europa.eu/eli/reg/2024/1183/oj}
}

@techreport{eidas_implementing_regulation,
  author      = {{European Commission}},
  title       = {{Draft implementing regulation on European Digital Identity Wallets certification}},
  institution = {European Commission},
  number      = {Ares(2024)5786790},
  year        = {2024},
  url         = {https://ec.europa.eu/info/law/better-regulation/have-your-say/initiatives/14337-European-Digital-Identity-Wallets-certification_en},
  note        = {Initiative 14337}
}

@misc{eidas_arf,
  author       = {{eu-digital-identity-wallet GitHub organization}},
  title        = {{EUDI Architecture and Reference Framework}},
  year         = {2025},
  howpublished = {GitHub release v2.4.0},
  url          = {https://github.com/eu-digital-identity-wallet/eudi-doc-architecture-and-reference-framework/releases/tag/v2.4.0},
  note         = {Accessed: 2025-09-05}
}

@misc{operation_triangulation,
  author = {Igor Kuznetsov and Valentin Pashkov and Leonid Bezvershenko and Georgy Kucherin},
  title  = {{Operation Triangulation: iOS devices targeted with previously unknown malware}},
  year   = {2023},
  howpublished = {Online},
  url    = {https://securelist.com/operation-triangulation/109842}
}

@inproceedings{alam_one_and_done,
  author     = {Monjur Alam and Haider Adnan Khan and Moumita Dey and Nishith Sinha and Robert Locke Callan and Alenka G. Zajic and Milos Prvulovic},
  title      = {One{\&}Done: {A} Single-Decryption {EM}-Based Attack on {OpenSSL}'s Constant-Time Blinded {RSA}},
  booktitle  = {27th {USENIX} Security Symposium ({USENIX} Security 2018)},
  pages      = {585--602},
  publisher  = {{USENIX} Association},
  year       = {2018},
  url        = {https://www.usenix.org/conference/usenixsecurity18/presentation/alam}
}

@misc{bhasin_aes_arm_cortex_a,
  author       = {Shivam Bhasin and Harishma Boyapally and Dirmanto Jap},
  title        = {{Reality Check on Side-Channels: Lessons learnt from breaking {AES} on an {ARM} Cortex A processor}},
  howpublished = {Cryptology {ePrint} Archive, Paper 2024/1381},
  year         = {2024},
  url          = {https://eprint.iacr.org/2024/1381}
}

@article{shepherd_physical_attacks_mobile_survey,
  author  = {Carlton Shepherd and Konstantinos Markantonakis and Nico van Heijningen and Driss Aboulkassimi and Cl{\'e}ment Gaine and Thibaut Heckmann and David Naccache},
  title   = {{Physical fault injection and side-channel attacks on mobile devices: {A} comprehensive analysis}},
  journal = {Computers \& Security},
  volume  = {111},
  pages   = {102471},
  year    = {2021},
  url     = {https://doi.org/10.1016/j.cose.2021.102471},
  doi     = {10.1016/J.COSE.2021.102471}
}

@inproceedings{cronin_charger_surfing,
  author     = {Patrick Cronin and Xing Gao and Chengmo Yang and Haining Wang},
  title      = {{Charger-Surfing: Exploiting a Power Line Side-Channel for Smartphone Information Leakage}},
  booktitle  = {30th {USENIX} Security Symposium ({USENIX} Security 2021)},
  pages      = {681--698},
  publisher  = {{USENIX} Association},
  year       = {2021},
  url        = {https://www.usenix.org/conference/usenixsecurity21/presentation/cronin}
}

@article{lisovets_iphone_bootloader,
  author  = {Oleksiy Lisovets and David Knichel and Thorben Moos and Amir Moradi},
  title   = {{Let's Take it Offline: Boosting Brute-Force Attacks on iPhone's User Authentication through {SCA}}},
  journal = {{IACR} Trans. Cryptogr. Hardw. Embed. Syst.},
  volume  = {2021},
  number  = {3},
  pages   = {496--519},
  year    = {2021},
  url     = {https://doi.org/10.46586/tches.v2021.i3.496-519},
  doi     = {10.46586/TCHES.V2021.I3.496-519}
}

@inproceedings{vasselle_bootloader,
  author     = {Aur{\'e}lien Vasselle and Philippe Maurine and Maxime Cozzi},
  title      = {{Breaking Mobile Firmware Encryption through Near-Field Side-Channel Analysis}},
  booktitle  = {ASHES@CCS 2019},
  pages      = {23--32},
  publisher  = {{ACM}},
  year       = {2019},
  url        = {https://doi.org/10.1145/3338508.3359571},
  doi        = {10.1145/3338508.3359571}
}

@inproceedings{longo_soc_it_to_em,
  author     = {Jake Longo and Elke De Mulder and Dan Page and Michael Tunstall},
  title      = {{SoC It to {EM}: ElectroMagnetic Side-Channel Attacks on a Complex System-on-Chip}},
  booktitle  = {Cryptographic Hardware and Embedded Systems ({CHES}) 2015},
  series     = {Lecture Notes in Computer Science},
  volume     = {9293},
  pages      = {620--640},
  publisher  = {Springer},
  year       = {2015},
  url        = {https://doi.org/10.1007/978-3-662-48324-4_31},
  doi        = {10.1007/978-3-662-48324-4_31}
}

@article{sayakkara_forensic_em,
  author  = {Asanka P. Sayakkara and Nhien-An Le-Khac},
  title   = {{Forensic Insights From Smartphones Through Electromagnetic Side-Channel Analysis}},
  journal = {{IEEE} Access},
  volume  = {9},
  pages   = {13237--13247},
  year    = {2021},
  url     = {https://doi.org/10.1109/ACCESS.2021.3051921},
  doi     = {10.1109/ACCESS.2021.3051921}
}

@inproceedings{genkin_ecdsa_mobile_nonintrusive,
  author     = {Daniel Genkin and Lev Pachmanov and Itamar Pipman and Eran Tromer and Yuval Yarom},
  title      = {{ECDSA} Key Extraction from Mobile Devices via Nonintrusive Physical Side Channels},
  booktitle  = {{ACM} {SIGSAC} Conference on Computer and Communications Security (CCS) 2016},
  pages      = {1626--1638},
  publisher  = {{ACM}},
  year       = {2016},
  url        = {https://doi.org/10.1145/2976749.2978353},
  doi        = {10.1145/2976749.2978353}
}

@inproceedings{goller_sca_smartphones_standard_radio,
  author     = {Gabriel Goller and Georg Sigl},
  title      = {{Side Channel Attacks on Smartphones and Embedded Devices Using Standard Radio Equipment}},
  booktitle  = {COSADE 2015},
  series     = {Lecture Notes in Computer Science},
  volume     = {9064},
  pages      = {255--270},
  publisher  = {Springer},
  year       = {2015},
  url        = {https://doi.org/10.1007/978-3-319-21476-4_17},
  doi        = {10.1007/978-3-319-21476-4_17}
}

@inproceedings{chen_infinity_gauntlet,
  author    = {Yu Chen and Yang Yu and Lidong Zhai},
  title     = {{InfinityGauntlet: brute-force attack on smartphone fingerprint authentication}},
  booktitle = {32nd {USENIX} Security Symposium ({USENIX} Security '23)},
  publisher = {{USENIX} Association},
  year      = {2023}
}

@inproceedings{tao_recovering_fingerprints,
  author    = {Tao Ni and Xiaokuan Zhang and Qingchuan Zhao},
  title     = {{Recovering Fingerprints from In-Display Fingerprint Sensors via Electromagnetic Side Channel}},
  booktitle = {{ACM} {SIGSAC} Conference on Computer and Communications Security (CCS) 2023},
  pages     = {253--267},
  publisher = {Association for Computing Machinery},
  year      = {2023},
  url       = {https://doi.org/10.1145/3576915.3623153},
  doi       = {10.1145/3576915.3623153}
}

@article{telleria_attack_potential_fingerprint,
  author  = {Ines Goicoechea-Telleria and Raul Sanchez-Reillo and Judith Liu-Jimenez and Ramon Blanco-Gonzalo},
  title   = {{Attack Potential Evaluation in Desktop and Smartphone Fingerprint Sensors: Can They Be Attacked by Anyone?}},
  journal = {Wireless Communications and Mobile Computing},
  volume  = {2018},
  number  = {1},
  pages   = {5609195},
  year    = {2018},
  url     = {https://onlinelibrary.wiley.com/doi/abs/10.1155/2018/5609195},
  doi     = {10.1155/2018/5609195}
}

@inproceedings{haas_apple_vs_ema,
  author    = {Gregor Haas and Aydin Aysu},
  title     = {{Apple vs. {EMA}: electromagnetic side channel attacks on Apple {CoreCrypto}}},
  booktitle = {59th {ACM}/{IEEE} Design Automation Conference (DAC)},
  pages     = {247--252},
  publisher = {Association for Computing Machinery},
  year      = {2022},
  url       = {https://doi.org/10.1145/3489517.3530437},
  doi       = {10.1145/3489517.3530437}
}

@article{navanesan_multiple_cpu_forensic,
  author  = {Lojenaa Navanesan and Kasun de Zoysa and Asanka P. Sayakkara},
  journal = {{IEEE} Access},
  title   = {{Impact of Multiple {CPU} Cores to the Forensic Insights Acquisition From Mobile Devices Using Electromagnetic Side-Channel Analysis}},
  year    = {2025},
  volume  = {13},
  pages   = {94953--94969},
  doi     = {10.1109/ACCESS.2025.3574340}
}

@inproceedings{tetsuya_ec_mult,
  author     = {Tetsuya Izu and Tsuyoshi Takagi},
  title      = {{A Fast Parallel Elliptic Curve Multiplication Resistant against Side Channel Attacks}},
  booktitle  = {Public Key Cryptography ({PKC}) 2002},
  series     = {Lecture Notes in Computer Science},
  volume     = {2274},
  pages      = {280--296},
  publisher  = {Springer},
  year       = {2002},
  url        = {https://doi.org/10.1007/3-540-45664-3_20},
  doi        = {10.1007/3-540-45664-3_20}
}

@inproceedings{nakano_preprocessing_smartphones,
  author     = {Yuto Nakano and Youssef Souissi and Robert Nguyen and Laurent Sauvage and Jean-Luc Danger and Sylvain Guilley and Shinsaku Kiyomoto and Yutaka Miyake},
  title      = {{A Pre-processing Composition for Secret Key Recovery on Android Smartphone}},
  booktitle  = {{IFIP} {WISTP} 2014},
  series     = {Lecture Notes in Computer Science},
  volume     = {8501},
  pages      = {76--91},
  publisher  = {Springer},
  year       = {2014},
  url        = {https://doi.org/10.1007/978-3-662-43826-8_6},
  doi        = {10.1007/978-3-662-43826-8_6}
}

@inproceedings{schneider_leakage_assessment,
  author     = {Tobias Schneider and Amir Moradi},
  title      = {{Leakage Assessment Methodology - {A} Clear Roadmap for Side-Channel Evaluations}},
  booktitle  = {{CHES} 2015},
  series     = {Lecture Notes in Computer Science},
  volume     = {9293},
  pages      = {495--513},
  publisher  = {Springer},
  year       = {2015},
  url        = {https://doi.org/10.1007/978-3-662-48324-4_25},
  doi        = {10.1007/978-3-662-48324-4_25}
}

@article{ascad,
  author  = {Ryad Benadjila and Emmanuel Prouff and R{\'e}mi Strullu and Eleonora Cagli and C{\'e}cile Dumas},
  title   = {{Deep learning for side-channel analysis and introduction to {ASCAD} database}},
  journal = {Journal of Cryptographic Engineering},
  volume  = {10},
  year    = {2020},
  doi     = {10.1007/s13389-019-00220-8}
}

@inproceedings{chari_template_attacks,
  author     = {Suresh Chari and Josyula R. Rao and Pankaj Rohatgi},
  title      = {{Template Attacks}},
  booktitle  = {{CHES} 2002},
  series     = {Lecture Notes in Computer Science},
  volume     = {2523},
  pages      = {13--28},
  publisher  = {Springer},
  year       = {2002},
  url        = {https://doi.org/10.1007/3-540-36400-5_3},
  doi        = {10.1007/3-540-36400-5_3}
}

@inproceedings{rechberger_practical_template_attacks,
  author     = {Christian Rechberger and Elisabeth Oswald},
  title      = {{Practical Template Attacks}},
  booktitle  = {{WISA} 2004},
  series     = {Lecture Notes in Computer Science},
  volume     = {3325},
  pages      = {440--456},
  publisher  = {Springer},
  year       = {2004},
  url        = {https://doi.org/10.1007/978-3-540-31815-6_35},
  doi        = {10.1007/978-3-540-31815-6_35}
}

@article{gohr_ches_challenge,
  author  = {Aron Gohr and Sven Jacob and Werner Schindler},
  title   = {{Efficient Solutions of the {CHES} 2018 {AES} Challenge Using Deep Residual Neural Networks and Knowledge Distillation on Adversarial Examples}},
  journal = {{IACR} Cryptology ePrint Archive},
  year    = {2020},
  url     = {https://eprint.iacr.org/2020/165}
}

@article{picek_sok_deep_learning_sca,
  author  = {Stjepan Picek and Guilherme Perin and Luca Mariot and Lichao Wu and Lejla Batina},
  title   = {{SoK: Deep Learning-based Physical Side-channel Analysis}},
  journal = {{ACM} Computing Surveys},
  volume  = {55},
  number  = {11},
  pages   = {227:1--227:35},
  year    = {2023},
  url     = {https://doi.org/10.1145/3569577},
  doi     = {10.1145/3569577}
}

@inproceedings{uellenbeck_unlock_patterns,
  author    = {Sebastian Uellenbeck and Markus D{\"u}rmuth and Christopher Wolf and Thorsten Holz},
  title     = {{Quantifying the Security of Graphical Passwords: The Case of Android Unlock Patterns}},
  booktitle = {{ACM} {SIGSAC} Conference on Computer and Communications Security (CCS) 2013},
  pages     = {161--172},
  publisher = {Association for Computing Machinery},
  year      = {2013},
  url       = {https://doi.org/10.1145/2508859.2516700},
  doi       = {10.1145/2508859.2516700}
}

@article{markert_unlock_pins,
  author    = {Philipp Markert and Daniel V. Bailey and Maximilian Golla and Markus D{\"u}rmuth and Adam J. Aviv},
  title     = {{On the Security of Smartphone Unlock {PINs}}},
  journal   = {{ACM} Transactions on Privacy and Security},
  volume    = {24},
  number    = {4},
  articleno = {30},
  pages     = {1--36},
  year      = {2021},
  url       = {https://doi.org/10.1145/3473040},
  doi       = {10.1145/3473040}
}

@inproceedings{leierzopf_adsdb,
  author    = {Ernst Leierzopf and Ren{\'e} Mayrhofer and Michael Roland and Wolfgang Studier and Lawrence Dean and Martin Seiffert and Florentin Putz and Lucas Becker and Daniel R. Thomas},
  title     = {{A Data-Driven Evaluation of the Current Security State of Android Devices}},
  booktitle = {{IEEE} Conference on Communications and Network Security (CNS) 2024},
  pages     = {1--9},
  year      = {2024},
  doi       = {10.1109/CNS62487.2024.10735682}
}

@techreport{nist_ecc,
  author      = {Lily Chen and Dustin Moody and Andrew Regenscheid and Angela Robinson and Karen Randall},
  title       = {{Recommendations for Discrete Logarithm-based Cryptography: Elliptic Curve Domain Parameters}},
  institution = {National Institute of Standards and Technology},
  number      = {NIST Special Publication (SP) 800-186},
  year        = {2023},
  doi         = {10.6028/NIST.SP.800-186}
}

@article{montgomery_ladder,
  author    = {Peter L. Montgomery},
  title     = {{Speeding the Pollard and elliptic curve methods of factorization}},
  journal   = {Mathematics of Computation},
  publisher = {American Mathematical Society (AMS)},
  year      = {1987},
  volume    = {48},
  number    = {177},
  pages     = {243--264},
  doi       = {10.1090/S0025-5718-1987-0866113-7}
}

@article{hintze_violin,
  author  = {Jerry L. Hintze and Ray D. Nelson},
  title   = {{Violin Plots: A Box Plot-Density Trace Synergism}},
  journal = {The American Statistician},
  volume  = {52},
  number  = {2},
  pages   = {181--184},
  year    = {1998},
  doi     = {10.1080/00031305.1998.10480559}
}

@inproceedings{boneh_hnp,
  author    = {Dan Boneh and Ramarathnam Venkatesan},
  title     = {{Hardness of Computing the Most Significant Bits of Secret Keys in Diffie-Hellman and Related Schemes}},
  booktitle = {Advances in Cryptology - {CRYPTO} '96},
  series    = {Lecture Notes in Computer Science},
  volume    = {1109},
  pages     = {129--142},
  publisher = {Springer},
  year      = {1996},
  doi       = {10.1007/3-540-68697-5_11}
}

@unpublished{bleichenbacher,
  author = {Daniel Bleichenbacher},
  title  = {{On the generation of one-time keys in {DL} signature schemes}},
  year   = {2000},
  note   = {Presentation at {IEEE} P1363 working group meeting}
}

@inproceedings{aranha_ladder_leak,
  author    = {Diego F. Aranha and Felipe Rodrigues Novaes and Akira Takahashi and Mehdi Tibouchi and Yuval Yarom},
  title     = {{LadderLeak: Breaking {ECDSA} with Less than One Bit of Nonce Leakage}},
  booktitle = {{ACM} {SIGSAC} Conference on Computer and Communications Security (CCS) 2020},
  pages     = {225--242},
  publisher = {Association for Computing Machinery},
  year      = {2020},
  url       = {https://doi.org/10.1145/3372297.3417268},
  doi       = {10.1145/3372297.3417268}
}

@inproceedings{demulder_bleichenbacher,
  author    = {Elke De Mulder and Michael Hutter and Mark E. Marson and Peter Pearson},
  title     = {{Using Bleichenbacher's Solution to the Hidden Number Problem to Attack Nonce Leaks in 384-Bit {ECDSA}}},
  booktitle = {{CHES} 2013},
  pages     = {435--452},
  publisher = {Springer},
  year      = {2013},
  url       = {https://doi.org/10.1007/978-3-642-40349-1_25},
  doi       = {10.1007/978-3-642-40349-1_25}
}

@inproceedings{albrecht_lattice_barrier,
  author    = {Martin R. Albrecht and Nadia Heninger},
  title     = {{On Bounded Distance Decoding with Predicate: Breaking the “Lattice Barrier” for the Hidden Number Problem}},
  booktitle = {Advances in Cryptology -- {EUROCRYPT} 2021},
  pages     = {528--558},
  publisher = {Springer},
  year      = {2021},
  url       = {https://doi.org/10.1007/978-3-030-77870-5_19},
  doi       = {10.1007/978-3-030-77870-5_19}
}

@article{xu_lattice_sieving,
  author  = {Luyao Xu and Zhengyi Dai and Baofeng Wu and Dongdai Lin},
  title   = {{Improved Attacks on (EC){DSA} with Nonce Leakage by Lattice Sieving with Predicate}},
  journal = {{IACR} Transactions on Cryptographic Hardware and Embedded Systems},
  volume  = {2023},
  number  = {2},
  pages   = {568--586},
  year    = {2023},
  url     = {https://tches.iacr.org/index.php/TCHES/article/view/10294},
  doi     = {10.46586/tches.v2023.i2.568-586}
}

@inproceedings{gao_lattice_sieving,
  author    = {Yiming Gao and Jinghui Wang and Honggang Hu and Binang He},
  title     = {{Attacking {ECDSA} with Nonce Leakage by Lattice Sieving: Bridging the Gap with Fourier Analysis-Based Attacks}},
  booktitle = {Advances in Cryptology -- {ASIACRYPT} 2024},
  pages     = {3--34},
  publisher = {Springer},
  year      = {2024},
  url       = {https://doi.org/10.1007/978-981-96-0944-4_1},
  doi       = {10.1007/978-981-96-0944-4_1}
}

@article{sun_guessing_bits,
  author  = {Chao Sun and Thomas Espitau and Mehdi Tibouchi and Masayuki Abe},
  title   = {{Guessing Bits: Improved Lattice Attacks on (EC){DSA} with Nonce Leakage}},
  journal = {{IACR} Transactions on Cryptographic Hardware and Embedded Systems},
  volume  = {2022},
  number  = {1},
  pages   = {391--413},
  year    = {2021},
  url     = {https://tosc.iacr.org/index.php/TCHES/article/view/9302},
  doi     = {10.46586/tches.v2022.i1.391-413}
}

@inproceedings{nascimento_ecc_cmov_sca,
  author    = {Erick Nascimento and {\L}ukasz Chmielewski and David Oswald and Peter Schwabe},
  title     = {{Attacking Embedded {ECC} Implementations Through cmov Side Channels}},
  booktitle = {Selected Areas in Cryptography -- {SAC} 2016},
  pages     = {99--119},
  publisher = {Springer},
  year      = {2017}
}

@inproceedings{nascimento_ecc_embedded_sca,
  author    = {Erick Nascimento and {\L}ukasz Chmielewski},
  title     = {{Applying Horizontal Clustering Side-Channel Attacks on Embedded {ECC} Implementations}},
  booktitle = {Smart Card Research and Advanced Applications},
  pages     = {213--231},
  publisher = {Springer},
  year      = {2018}
}

@inproceedings{genkin_ecdh_sca_pc,
  author    = {Daniel Genkin and Lev Pachmanov and Itamar Pipman and Eran Tromer},
  title     = {{ECDH} Key-Extraction via Low-Bandwidth Electromagnetic Attacks on {PCs}},
  booktitle = {Topics in Cryptology - {CT-RSA} 2016},
  pages     = {219--235},
  publisher = {Springer},
  year      = {2016}
}

@inproceedings{belgarric_ecdsa_android,
  author    = {Pierre Belgarric and Pierre-Alain Fouque and Gilles Macario-Rat and Mehdi Tibouchi},
  title     = {{Side-Channel Analysis of Weierstrass and Koblitz Curve {ECDSA} on Android Smartphones}},
  booktitle = {Topics in Cryptology - {CT-RSA} 2016},
  pages     = {236--252},
  publisher = {Springer},
  year      = {2016},
  url       = {https://doi.org/10.1007/978-3-319-29485-8_14},
  doi       = {10.1007/978-3-319-29485-8_14}
}

@article{batina_online_ta,
  author  = {Lejla Batina and {\L}ukasz Chmielewski and Louiza Papachristodoulou and Peter Schwabe and Michael Tunstall},
  title   = {{Online template attacks}},
  journal = {Journal of Cryptographic Engineering},
  volume  = {9},
  number  = {1},
  pages   = {21--36},
  year    = {2019},
  doi     = {10.1007/s13389-017-0171-8},
  url     = {https://doi.org/10.1007/s13389-017-0171-8}
}

@inproceedings{dugardin_dismantling_ecc_ta,
  author    = {Margaux Dugardin and Louiza Papachristodoulou and Zakaria Najm and Lejla Batina and Jean-Luc Danger and Sylvain Guilley},
  title     = {{Dismantling Real-World {ECC} with Horizontal and Vertical Template Attacks}},
  booktitle = {Constructive Side-Channel Analysis and Secure Design},
  pages     = {88--108},
  publisher = {Springer},
  year      = {2016}
}

@inproceedings{roelofs_online_ta_other_side,
  author    = {Niels Roelofs and Niels Samwel and Lejla Batina and Joan Daemen},
  title     = {{Online Template Attack on {ECDSA}: Extracting Keys via the Other Side}},
  booktitle = {{AFRICACRYPT} 2020},
  pages     = {323--336},
  publisher = {Springer},
  year      = {2020},
  url       = {https://doi.org/10.1007/978-3-030-51938-4_16},
  doi       = {10.1007/978-3-030-51938-4_16}
}

@inproceedings{weissbart_one_trace,
  author    = {Leo Weissbart and Stjepan Picek and Lejla Batina},
  title     = {{One Trace Is All It Takes: Machine Learning-Based Side-Channel Attack on {EdDSA}}},
  booktitle = {{SPACE} 2019},
  series    = {Lecture Notes in Computer Science},
  volume    = {11947},
  pages     = {86--105},
  publisher = {Springer},
  year      = {2019},
  url       = {https://doi.org/10.1007/978-3-030-35869-3_8},
  doi       = {10.1007/978-3-030-35869-3_8}
}

@inproceedings{medwed_ta_ecdsa,
  author    = {Marcel Medwed and Elisabeth Oswald},
  title     = {{Template Attacks on {ECDSA}}},
  booktitle = {Information Security Applications},
  pages     = {14--27},
  publisher = {Springer},
  year      = {2009}
}

@article{batina_sca_secure_ecc_sok,
  author  = {Lejla Batina and Lukasz Chmielewski and Bj{\"o}rn Haase and Niels Samwel and Peter Schwabe},
  title   = {{SoK: {SCA}-secure {ECC} in Software - Mission Impossible?}},
  journal = {{IACR} Transactions on Cryptographic Hardware and Embedded Systems},
  volume  = {2023},
  number  = {1},
  pages   = {557--589},
  year    = {2023},
  url     = {https://doi.org/10.46586/tches.v2023.i1.557-589},
  doi     = {10.46586/TCHES.V2023.I1.557-589}
}

@article{jin_ecdsa_collision,
  author  = {Sunghyun Jin and Sangyub Lee and Sung Min Cho and HeeSeok Kim and Seokhie Hong},
  title   = {{Novel Key Recovery Attack on Secure {ECDSA} Implementation by Exploiting Collisions between Unknown Entries}},
  journal = {{IACR} Transactions on Cryptographic Hardware and Embedded Systems},
  volume  = {2021},
  number  = {4},
  pages   = {1--26},
  year    = {2021},
  url     = {https://tches.iacr.org/index.php/TCHES/article/view/9058},
  doi     = {10.46586/tches.v2021.i4.1-26}
}

@inproceedings{kocher_sca,
  author    = {Paul C. Kocher},
  title     = {{Timing Attacks on Implementations of Diffie-Hellman, {RSA}, {DSS}, and Other Systems}},
  booktitle = {Advances in Cryptology --- {CRYPTO} '96},
  pages     = {104--113},
  publisher = {Springer},
  year      = {1996}
}

@article{kocher_dpa,
  author  = {Paul Kocher and Joshua Jaffe and Benjamin Jun and Pankaj Rohatgi},
  title   = {{Introduction to Differential Power Analysis}},
  journal = {Journal of Cryptographic Engineering},
  volume  = {1},
  number  = {1},
  pages   = {5--27},
  year    = {2011},
  doi     = {10.1007/s13389-011-0006-y},
  url     = {https://doi.org/10.1007/s13389-011-0006-y}
}

@inproceedings{kocher_spectre,
  author    = {Paul Kocher and Jann Horn and Anders Fogh and Daniel Genkin and Daniel Gruss and Werner Haas and Mike Hamburg and Moritz Lipp and Stefan Mangard and Thomas Prescher and Michael Schwarz and Yuval Yarom},
  title     = {{Spectre Attacks: Exploiting Speculative Execution}},
  booktitle = {{IEEE} Symposium on Security and Privacy ({SP}) 2019},
  pages     = {1--19},
  year      = {2019},
  doi       = {10.1109/SP.2019.00002}
}

@inproceedings{agrawal_em_sca,
  author    = {Dakshi Agrawal and Bruce Archambeault and Josyula R. Rao and Pankaj Rohatgi},
  title     = {{The {EM} Side---Channel(s)}},
  booktitle = {{CHES} 2002},
  pages     = {29--45},
  publisher = {Springer},
  year      = {2003}
}

@techreport{tuev_qualcomm_tee,
  author      = {Denise Cater},
  title       = {{Certification Report: Qualcomm\textsuperscript{\textregistered} Trusted Execution Environment ({TEE}) v5.8 on Qualcomm\textsuperscript{\textregistered} Snapdragon\texttrademark{} 865}},
  institution = {T{\"U}V Rheinland Nederland B.V.},
  year        = {2021},
  month       = {aug},
  url         = {https://www.commoncriteriaportal.org/nfs/ccpfiles/files/epfiles/NSCIB-CC-0244671-CR-1.0.pdf}
}

@techreport{trustcb_titan_certificate,
  author      = {TrustCB B.V.},
  title       = {Certificate: {H1D3} Secure Microcontroller with Crypto Library v1.3.10},
  institution = {TrustCB B.V.},
  year        = {2023},
  month       = {oct},
  url         = {https://www.commoncriteriaportal.org/nfs/ccpfiles/files/epfiles/NSCIB-CC-2300073-02-CR.pdf}
}

@inproceedings{bernstein_highspeed_signatures,
  author    = {Daniel J. Bernstein and Niels Duif and Tanja Lange and Peter Schwabe and Bo-Yin Yang},
  title     = {{High-speed high-security Signatures}},
  booktitle = {{CHES} 2011},
  pages     = {124--142},
  publisher = {Springer},
  year      = {2011}
}

@article{zheng_spoofing_fa,
  author  = {Zheng Zheng and Qian Wang and Cong Wang},
  title   = {{Spoofing Attacks and Anti-Spoofing Methods for Face Authentication Over Smartphones}},
  journal = {{IEEE} Communications Magazine},
  volume  = {61},
  number  = {12},
  pages   = {213--219},
  year    = {2023},
  doi     = {10.1109/MCOM.012.2200794}
}

@inproceedings{vedros_code_to_em,
  author    = {Kurt A. Vedros and Constantinos Kolias and Daniel Barbara and Robert C. Ivans},
  title     = {{From Code to {EM} Signals: A Generative Approach to Side Channel Analysis-based Anomaly Detection}},
  booktitle = {International Conference on Availability, Reliability and Security (ARES) 2024},
  publisher = {Association for Computing Machinery},
  year      = {2024},
  url       = {https://doi.org/10.1145/3664476.3664520},
  doi       = {10.1145/3664476.3664520}
}

@article{yilmaz_capacity_covert_channel_instructions,
  author  = {Baki Berkay Yilmaz and Robert L. Callan and Milos Prvulovic and Alenka Zaji{\'c}},
  title   = {{Capacity of the {EM} Covert/Side-Channel Created by the Execution of Instructions in a Processor}},
  journal = {{IEEE} Transactions on Information Forensics and Security},
  volume  = {13},
  number  = {3},
  pages   = {605--620},
  year    = {2018},
  doi     = {10.1109/TIFS.2017.2762826}
}

@inproceedings{callan_fase,
  author    = {Robert Callan and Alenka Zaji{\'c} and Milos Prvulovic},
  title     = {{FASE: finding amplitude-modulated side-channel emanations}},
  booktitle = {International Symposium on Computer Architecture ({ISCA}) 2015},
  pages     = {592--603},
  publisher = {Association for Computing Machinery},
  year      = {2015},
  url       = {https://doi.org/10.1145/2749469.2750394},
  doi       = {10.1145/2749469.2750394}
}

@article{prvulovic_finding_fm_am,
  author  = {Milos Prvulovic and Alenka Zaji{\'c} and Robert L. Callan and Christopher J. Wang},
  title   = {{A Method for Finding Frequency-Modulated and Amplitude-Modulated Electromagnetic Emanations in Computer Systems}},
  journal = {{IEEE} Transactions on Electromagnetic Compatibility},
  volume  = {59},
  number  = {1},
  pages   = {34--42},
  year    = {2017},
  doi     = {10.1109/TEMC.2016.2603847}
}

@inproceedings{wang_finding_carriers,
  author    = {Christopher Wang and Robert Callan and Alenka Zaji{\'c} and Milos Prvulovic},
  title     = {{An algorithm for finding carriers of amplitude-modulated electromagnetic emanations in computer systems}},
  booktitle = {10th European Conference on Antennas and Propagation (EuCAP) 2016},
  year      = {2016},
  doi       = {10.1109/EuCAP.2016.7481633}
}

@techreport{snapdragon750,
  title       = {{Qualcomm Snapdragon 750G 5G Mobile Platform Product Brief}},
  institution = {Qualcomm Technologies, Inc.},
  year        = {2020},
  url         = {https://www.qualcomm.com/content/dam/qcomm-martech/dm-assets/documents/snapdragon_750g_5g_mobile_platform_product_brief_0.pdf},
  note        = {Accessed: 2025-08}
}

@inproceedings{itoh_countermeasure_address_dpa,
  author    = {Kouichi Itoh and Tetsuya Izu and Masahiko Takenaka},
  title     = {{A Practical Countermeasure against Address-Bit Differential Power Analysis}},
  booktitle = {{CHES} 2003},
  pages     = {382--396},
  publisher = {Springer},
  year      = {2003}
}

@inproceedings{heyszl_localem,
  author    = {Johann Heyszl and Stefan Mangard and Benedikt Heinz and Frederic Stumpf and Georg Sigl},
  title     = {{Localized Electromagnetic Analysis of Cryptographic Implementations}},
  booktitle = {Topics in Cryptology -- {CT-RSA} 2012},
  pages     = {231--244},
  publisher = {Springer},
  year      = {2012}
}

@inproceedings{coron_resistance_dpa_ecc,
  author    = {Jean-S{\'e}bastien Coron},
  title     = {{Resistance Against Differential Power Analysis For Elliptic Curve Cryptosystems}},
  booktitle = {{CHES} 1999},
  pages     = {292--302},
  publisher = {Springer},
  year      = {1999}
}

@inproceedings{suvanka_telegram,
  author    = {Tom{\'a}{\v{s}} Su{\v{s}}{\'a}nka and Josef Koke{\v{s}}},
  title     = {{Security Analysis of the Telegram {IM}}},
  booktitle = {Reversing and Offensive-Oriented Trends Symposium (ROOTS) 2017},
  publisher = {Association for Computing Machinery},
  year      = {2017},
  url       = {https://doi.org/10.1145/3150376.3150382},
  doi       = {10.1145/3150376.3150382}
}

@inproceedings{aviv_smudge_attacks,
  author    = {Adam J. Aviv and Katherine Gibson and Evan Mossop and Matt Blaze and Jonathan M. Smith},
  title     = {{Smudge attacks on smartphone touch screens}},
  booktitle = {{USENIX} Workshop on Offensive Technologies (WOOT) 2010},
  publisher = {{USENIX} Association},
  year      = {2010}
}

@techreport{cc_eal,
  title       = {{Common Criteria for Information Technology Security Evaluation, Part 5}},
  institution = {Common Criteria Development Board},
  year        = {2022},
  type        = {CC:2022, Release 1},
  url         = {https://www.commoncriteriaportal.org/files/ccfiles/CC2022PART5R1.pdf},
  note        = {Accessed: 2025-09-11}
}

@misc{android_keystore,
  author       = {{Android Developers}},
  title        = {{Android Keystore System}},
  year         = {2025},
  howpublished = {Online},
  url          = {https://developer.android.com/privacy-and-security/keystore},
  note         = {Accessed: 2025-09-16}
}

@misc{google_auto_reboot,
  author       = {Google},
  title        = {{Google System Services Release Notes}},
  year         = {2025},
  howpublished = {Online},
  url          = {https://support.google.com/product-documentation/answer/14343500},
  note         = {Accessed: 2025-09-16}
}

@misc{google_pixel_security,
  author       = {Google},
  title        = {{Pixel security}},
  year         = {2025},
  howpublished = {Online},
  url          = {https://safety.google/pixel/},
  note         = {Accessed: 2025-09-16}
}

@inproceedings{wang_pixnapping,
  author = {Alan Wang and Pranav Gopalkrishnan and Yingchen Wang and Christopher W. Fletcher and Hovav Shacham and David Kohlbrenner and Riccardo Paccagnella},
  title = {Pixnapping: Bringing Pixel Stealing out of the Stone Age},
  booktitle = {Proceedings of the ACM Conference on Computer and Communications Security (CCS)},
  year = {2025}
}

@techreport{eucc_2025,
  title        = {{EUCC Scheme State-of-the-Art Document: Application of Attack Potential to Smartcards and Similar Devices}},
  author       = {{European Union Agency for Cybersecurity (ENISA)}},
  year         = {2025},
  month        = feb,
  institution  = {European Union Agency for Cybersecurity (ENISA)},
  type         = {Version 2},
  note         = {Endorsed for publication on 11/03/2025 by the European Cybersecurity Certification Group (ECCG)},
  url          = {https://certification.enisa.europa.eu/index_en}
}

\appendices
\crefalias{section}{appendix}
\section{Acknowledgment}

This research was supported by the German Federal Office for Information Security (BSI) and the Bavarian Ministry of Economic Affairs, Regional Development and Energy in the context of the project Trusted Electronics Bavaria (TrEB).

%

\section{Disclosure}
\label{appendix:disclosure}

We adhered to a coordinated disclosure process for this work.
In October 2025, we notified the OpenSSL and libgcrypt security teams about the demonstrated side‑channel leakage in their \ac{ecdsa} implementations and provided our assessment of existing countermeasures.
Please note that while both libraries may choose to integrate countermeasures against physical side-channel attacks, they do not consider such attacks in their threat models\footnote{\url{https://openssl-library.org/policies/general/security-policy/}}\footnote{\url{https://www.gnupg.org/documentation/security.html}}.

At the same time we informed Broadcom and the Raspberry Pi Foundation as well as Qualcomm and Fairphone about our empirical findings on the exemplary Raspberry Pi 4 (BCM2711 \ac{soc}) and Fairphone 4 (Snapdragon 750G 5G \ac{soc}) targets, including our threat models, required lab setup, and the distinction between invasive and non‑invasive measurement scenarios.

None of the contacted entities voiced concerns about this publication.

\section{Raspberry Pi 4 and the BCM2711 \ac{soc}}
\label{appendix:rpi4}

\begin{figure}
\centering
	\begin{subfigure}[t]{0.49\linewidth}
	\centering
	\includegraphics[width=\textwidth]{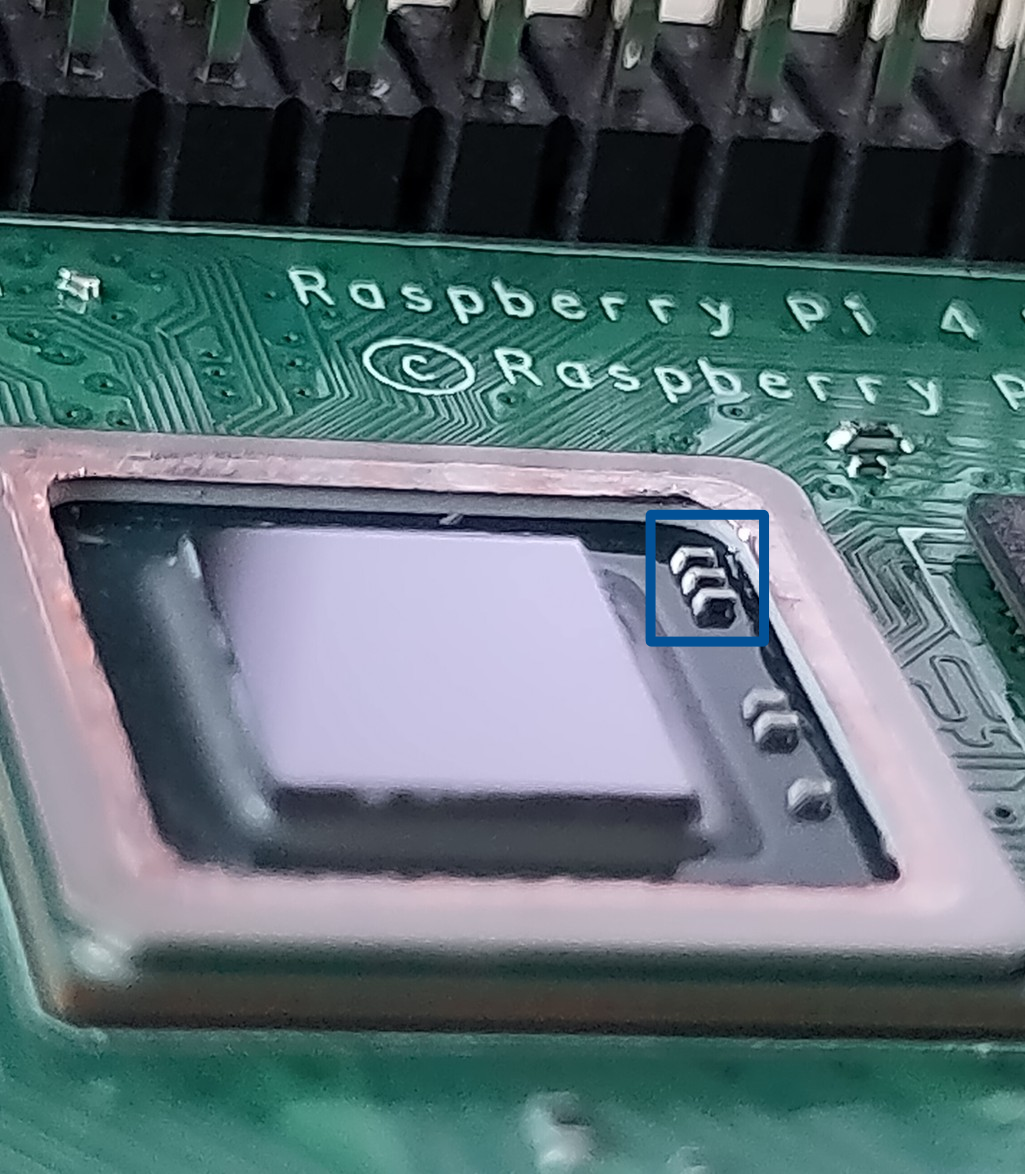}
	\caption{The Raspberry Pi 4's BCM2711 chip with removed metal lid. On the right side of the die, six capacitors are visible.}
	\label{fig:rpi4chip}
\end{subfigure}
\hfill
	\begin{subfigure}[t]{0.49\linewidth}
	\centering
	\includegraphics[width=\textwidth]{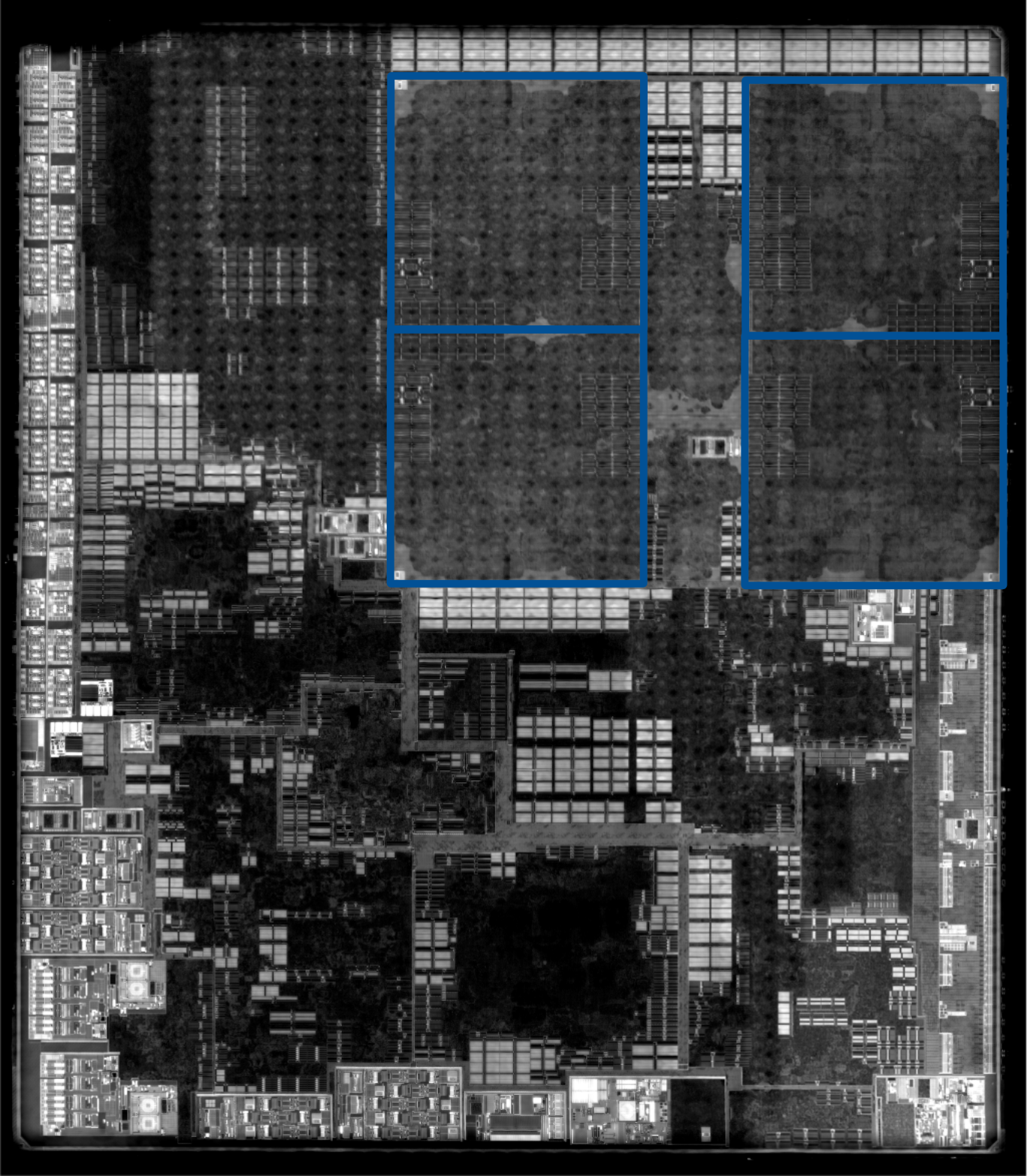}
	\caption{\ac{ir} die-shot of the BCM2711 \ac{soc}. The four ARM Cortex-A72 cores are outlined with blue rectangles.}
	\label{fig:rpi4dieshot}
\end{subfigure}
\caption{Raspberry Pi 4 Model B Target}
\end{figure}

\cref{fig:rpi4chip} shows the Broadcom BCM2711 \ac{soc} with removed metal lid.
We recorded an \ac{ir} die-shot to determine the location of the four \acp{cpu} for localized measurements on the die (\cref{fig:rpi4dieshot}).
The die-shot was acquired with a WIDY SWIR 640V-S camera from New Imaging Technologies.

\section{Montgomery ladder step in OpenSSL}
\label{appendix:montladderstep}

OpenSSL uses \cref{alg:montladder} for the step operation of the Montgomery ladder.
Each odd line in \cref{alg:montladder} contains finite-field multiplication and square operations.
The even lines contain addition, subtraction or shift operations (used to implement the multiplication with a power of two), which are comparatively short.
The operations can be clearly mapped to the \textit{5-2-1-2-3-1-3-3} pattern of blocks in the bandpass-filtered signal and adjacent peaks in the bandpass + absolute + sliding-median filtered signal of \cref{fig:rpi4osslgcry}.
Each multiplication/square operation leads to a block/peak and the operations on the even lines make up the breaks between these blocks/peaks.
For example, line \num{1} corresponds to the \num{5} adjacent peaks at the beginning of the trace in \cref{fig:rpi4osslgcry}.

\begin{algorithm}
\caption{Montgomery ladder step in OpenSSL with differential addition-and-doubling formulas from~\cite{tetsuya_ec_mult}.}
\label{alg:montladder}
\hspace*{\algorithmicindent} \textbf{Input:}\\Weierstrass equation: $y^2 \equiv x^3 + ax + b \pmod p$;\\Input points in projective coordinates: $s=(X_1:Y_1:Z_1)$, $r=(X_2:Y_2:Z_2)$;\\Generator point in affine coordinates: $p=(x,y)$.\\
\hspace*{\algorithmicindent} \textbf{Output:}\\Result in $s$, $r$ as projective coordinates: $s \gets r + s$, $r \gets 2r$.
\begin{algorithmic}[1]
\State $t_6 \gets X_2 \cdot X_1$; $t_0 \gets Z_2 \cdot Z_1$; $t_4 \gets X_2 \cdot Z_1$; $t_3 \gets Z_2 \cdot X_1$; $t_5 \gets a \cdot t_0$
\State $t_4 \gets t_6 + t_5$; $t_6 \gets t_3 + t_4$
\State $t_5 \gets t_6 \cdot t_5$; $t_0 \gets t_0^2$
\State $t_2 \gets b \ll 2$
\State $t_0 \gets t_2 \cdot t_0$
\State $t_5 \gets t_5 \ll 1$; $t_3 \gets t_4 - t_3$
\State $Z_1 \gets t_3^2$; $t_4 \gets Z_1 \cdot x$
\State $t_0 \gets t_0 + t_5$; $X_1 \gets t_0 - t_4$
\State $t_4 \gets X_2^2$; $t_5 \gets Z_2^2$; $t_6 \gets a \cdot t_5$
\State $t_1 \gets X_2 + Z_2$
\State $t_1 \gets t_1^2$
\State $t_1 \gets t_1 - t_4$; $t_1 \gets t_1 - t_5$; $t_3 \gets t_4 - t_6$
\State $t_3 \gets t_3^2$; $t_0 \gets t_5 \cdot t_1$; $t_0 \gets t_2 \cdot t_0$
\State $X_2 \gets t_3 - t_0$; $t_3 \gets t_4 + t_6$
\State $t_4 \gets t_5^2$; $t_4 \gets t_4 \cdot t_2$; $t_1 \gets t_1 \cdot t_3$
\State $t_1 \gets t_1 \ll 1$; $Z_2 \gets t_4 + t_1$
\end{algorithmic}
\end{algorithm}

\section{Evaluation of classifiers for swap conditions}
\label{appendix:classifiers}

\begin{figure}
	\centering
	\includegraphics[width=\linewidth]{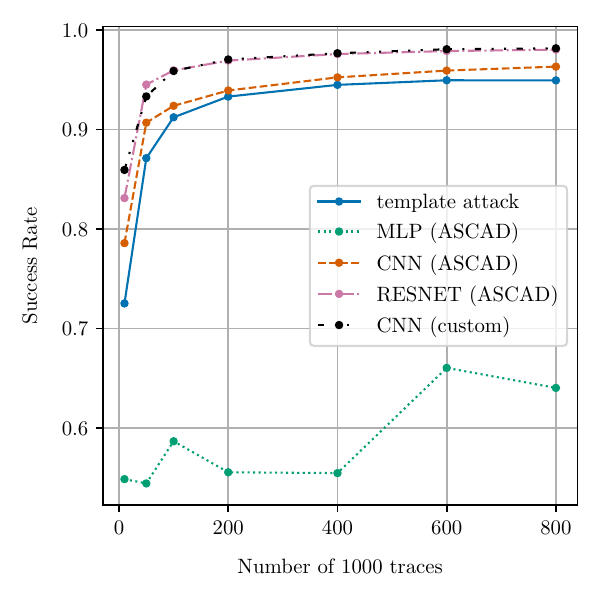}
        \caption{Success rate of various classifiers in relation to number of training traces.}
	\label{fig:rpi4classifiers}
\end{figure}

\cref{fig:rpi4classifiers} shows the accuracy of multiple classifiers.
The \ac{ascad} networks were taken from commit id $410b92b$\footnote{\url{https://github.com/ANSSI-FR/ASCAD/tree/410b92bab3bc69502b43c5cc9ccdec74794870be}}.
\ac{ascad} includes $CNN_1$ and $CNN_2$, which differ by the fact that $CNN_2$ uses a stride length of two for the convolution layer in the first block.
Our results are for $CNN_1$.
Since those networks were designed for predicting a byte within the \ac{aes} state, we also conducted a short hyper parameter search, which led to the architecture described in \cref{tab:cnnarch}.
At a success rate of \SI{98.14}{\percent}, this custom \ac{cnn} performs marginally better than the residual neural network taken from the \ac{ascad} project.
At an accuracy of \SI{94.92}{\percent}, the Gaussian template attack \cite{chari_template_attacks, rechberger_practical_template_attacks} is outperformed by all networks but the \ac{mlp}.

\begin{table}
\centering
\caption{Architecture of the swap condition classifier \ac{cnn}.}
\label{tab:cnnarch}
\begin{tabular}{@{}ll@{}}
\toprule
\textbf{Layer}               & \textbf{Description} \\
\midrule
Input                        & Input of shape $(1000, 1)$ \\
Batch Normalization	     & \\
\midrule
Conv1D 		             & 32 filters, kernel size 90, ReLU \\
Conv1D		             & 32 filters, kernel size 90, ReLU \\
Average Pooling              & Pool size=2, strides=2 \\
\midrule
Conv1D		             & 64 filters, kernel size 90, ReLU \\
Conv1D 		             & 64 filters, kernel size 90, ReLU \\
Average Pooling              & Pool size=2, strides=2 \\
\midrule
\multicolumn{2}{c}{... (3 more blocks with doubling of filters) ...} \\
\midrule
Flatten			    & \\
Dense                       & Fully connected, 4096 neurons, ReLU \\
Dense (Output)              & Fully connected, 2 neurons, softmax \\
\bottomrule
\end{tabular}
\end{table}

\section{Snapdragon 750G 5G \ac{soc}}
\label{appendix:snapdragon750}

\begin{figure}
	\centering
	\includegraphics[width=\linewidth]{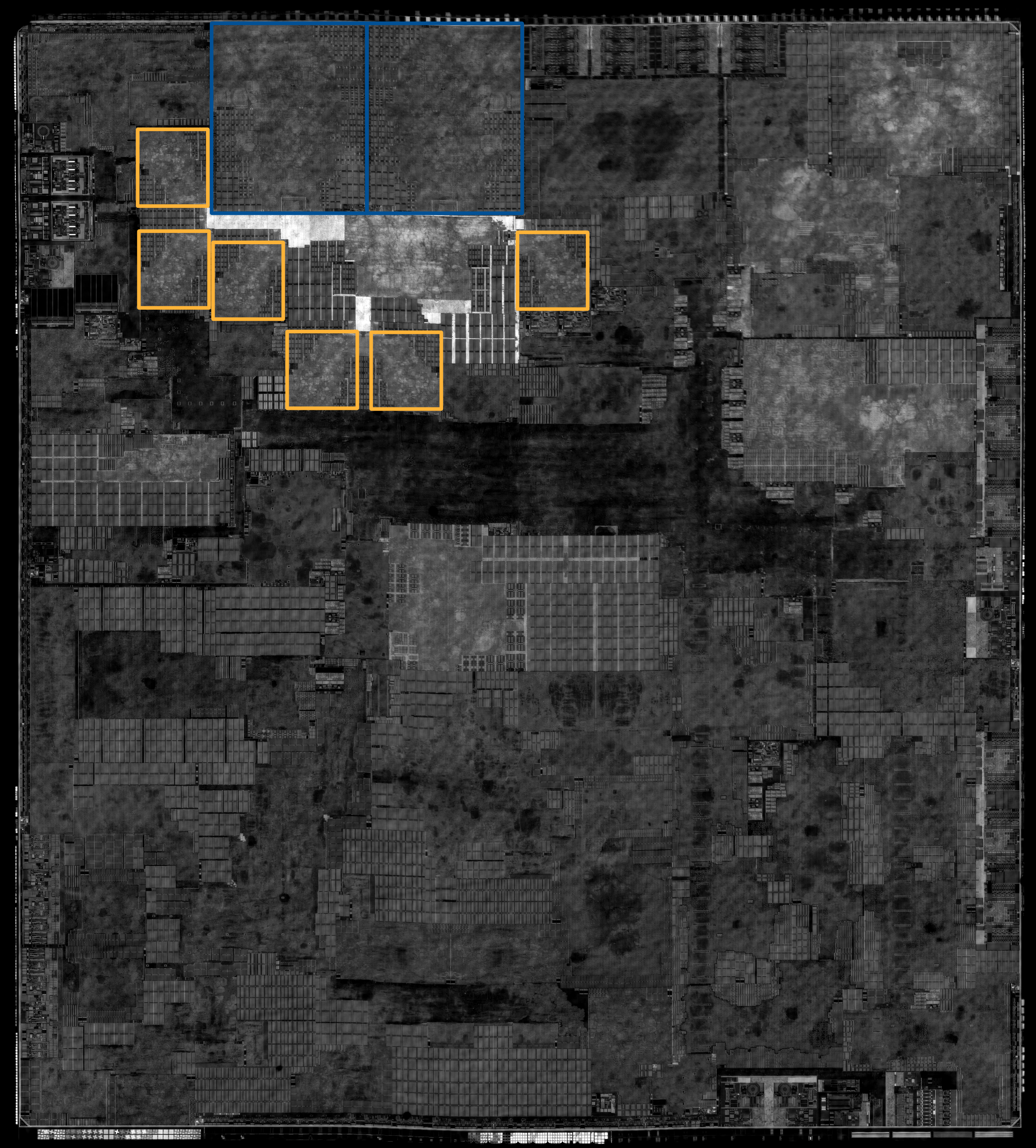}
	\caption{\ac{ir} die-shot of the Snapdragon 750G 5G \ac{soc}. The two ARM Cortex-A77 cores are outlined in blue, the six ARM Cortex-A55 cores are marked with yellow rectangles.}
	\label{fig:fp4dieshot}
\end{figure}

\cref{fig:fp4dieshot} shows an \ac{ir} die-shot of the \ac{soc}.
The \ac{soc} includes a Kryo 570 octa-core \ac{cpu}~\cite{snapdragon750}.
Due to their distinct structures, the two A77 and six A55 cores can be identified.
By installing postmarketOS and using the \textit{cpufreq-info} tool, we obtained the supported frequencies.
The Cortex-A55 \acp{cpu} can be clocked at the following frequencies:

\begin{itemize}
\begin{multicols}{2}
		\item{\SI{300}{\mega\hertz}}
		\item{\SI{576}{\mega\hertz}}
		\item{\SI{768}{\mega\hertz}}
		\item{\SI{1.02}{\giga\hertz}}
		\item{\SI{1.25}{\giga\hertz}}
		\item{\SI{1.32}{\giga\hertz}}
		\item{\SI{1.52}{\giga\hertz}}
		\item{\SI{1.61}{\giga\hertz}}
		\item{\SI{1.71}{\giga\hertz}}
		\item{\SI{1.80}{\giga\hertz}}
\end{multicols}
\end{itemize}

The Cortex-A77 \acp{cpu} can be clocked at:

\begin{itemize}
\begin{multicols}{2}
		\item \SI{300}{\mega\hertz}
		\item \SI{787}{\mega\hertz}
		\item \SI{979}{\mega\hertz}
		\item \SI{1.04}{\giga\hertz}
		\item \SI{1.25}{\giga\hertz}
		\item \SI{1.40}{\giga\hertz}
		\item \SI{1.56}{\giga\hertz}
		\item \SI{1.77}{\giga\hertz}
		\item \SI{1.90}{\giga\hertz}
		\item \SI{2.07}{\giga\hertz}
		\item \SI{2.13}{\giga\hertz}
		\item \SI{2.21}{\giga\hertz}
\end{multicols}
\end{itemize}

\end{document}